\documentclass{aastex}
\usepackage{emulateapj5,psfig}

\newcommand{\kms}{km s$^{-1}$}
\newcommand{\msun}{$M_{\sun}$}
\newcommand{\hubu}{km s$^{-1}$ Mpc$^{-1}$}

\journalinfo{Accepted for publication in AJ}

\shortauthors{Tran et al.}
\shorttitle{Dust in Early-Type Galaxies}

\begin{document}


\title{Dusty Nuclear Disks and Filaments in Early Type Galaxies\footnote{
Based on observations with the NASA/ESA {\it Hubble Space Telescope}, obtained at the 
Space Telescope Science Institute, which is operated by the Association of Universities for Research 
in Astronomy, Inc. under NASA contract No. NAS5-26555.}}


\author{H. D. Tran, Z. Tsvetanov, H. C. Ford and J. Davies}
\affil{Department of Physics and Astronomy, Johns Hopkins University,
    Baltimore, MD 21218}

\author{W. Jaffe}
\affil{Leiden Observatory, P.O. Box 9513, 2300 RA Leiden, The Netherlands}

\and

\author{F. C. van den Bosch\altaffilmark{1}, A. Rest}
\affil{Department of Astronomy, University of Washington, Seattle, WA 98195}

\altaffiltext{1}{Present address: MPI fur Astrophysik, Karl Schwarzschild Str. 1, Postfach 1317,
85741 Garching, Germany.}

\begin{abstract}
We examine the dust properties of a nearby distance-limited sample of early type
galaxies using the WFPC2 of the $Hubble~Space~Telescope$. Dust is detected in 29 out of
67 galaxies (43\%), including 12 with small nuclear dusty disks. 
In a separate sample of 40 galaxies biased for the detection of dust by virtue of their 
detection in the $IRAS$ 100 $\mu$m band, dust is found in $\sim$ 78\% of the galaxies, 15 of which
contain dusty disks. 
In those galaxies with detectable dust, the apparent mass of the dust correlates
with radio and far infrared luminosity, becoming more significant for systems with
filamentary dust. 
A majority of $IRAS$ and radio detections are also associated with dusty galaxies rather than 
dustless galaxies.
This indicates that thermal emission from clumpy, filamentary dust is 
the main source of the far-IR radiation in early type galaxies. 
Dust in small disk-like morphology tends to be well aligned with the major axis of the
host galaxies, while filamentary dust appears to be more randomly distributed with no
preference for alignment with any major galactic structure. This suggests that, if the dusty
disks and filaments have a common origin, the dust originates externally and requires time 
to dynamically relax and settle in the galaxy potential in the form of compact disks.  
More galaxies with visible dust than without dust display emission lines, indicative of 
ionized gas, although such nuclear activity does not show a preference 
for dusty disk over filamentary dust.
There appears to be a weak relationship between the mass of the dusty
disks and central velocity dispersion of the galaxy, suggesting a connection with
a similar recently recognized relationship between the latter and the black hole mass.

\end{abstract}

\keywords{galaxies: elliptical and lenticular --- galaxies: ISM --- ISM: dust, extinction}

\section{Introduction} \label{intro}
Elliptical galaxies have often been viewed as old, uniform
systems, with little gas, dust and activity. Through the years, ground-based and especially high
spatial resolution imaging studies with the $Hubble~Space~Telescope$ ($HST$) have rapidly 
changed that view \citep{sg85,vv88,f91,gou94,f94,j94,vdb94,dkfr95,l95,fab97,fer99,tom00}. 
Based on these surveys, early-type galaxies are now known to commonly contain large amount 
of dust in various forms and sometimes complex structures. 
Of particular interest are the small ($r \sim 1$\arcsec, 200 pc) dusty nuclear disks that 
have been found in the centers of elliptical galaxies 
(see review by Ford et al. 1997). Their discovery and the ionized gas disks, associated with the
dust, opened up a new means of efficiently measuring the central mass potentials, possibly of 
supermassive black holes (BH), residing in the centers of galaxies 
\citep{h94,ffj96,mac97,bow98,vv98,ff99,vk00}. 

In this paper, we conduct a survey of two large samples of elliptical 
galaxies to study their dust properties and how they are related to other global
properties of the parent galaxies as a whole. 
The main sample consists of E or S0 galaxies from the Lyon/Meudon Extragalactic 
Database (LEDA), selected to be nearby with $v <$ 3400 \kms, and lying
at galactic latitude $>$ 20\arcdeg~to minimize Galactic extinction. 
Details of the sample selection and morphological and photometric properties of the 
galaxies are described in \citet{rbos00} (hereafter Paper I).
A total of 130 galaxies meet our criteria, out of which 67 galaxies were observed
by $HST$ using WFPC2 in snapshot mode with the F702W filter ($R$ band). 
We shall refer to this sample as the ``snapshot'' sample in the rest of the paper.

In order to better assess the results derived from the snapshot sample, we 
also discuss where relevant a sample of galaxies selected for their 100 $\mu$m $IRAS$ emission.
The rational is that these galaxies are selected for their higher far-IR (60 and 100 $\mu$m) emission,
and thus are probably more likely to contain large amount of dust. 
The selection criteria are similar to those of the snapshot sample except that the galaxies are 
drawn from the $HST$ archive, and with the additional criterion that a 100 $\mu$m $IRAS$ detection 
exist with $\gtrsim 3\sigma$ significance. 
A total of 40 galaxies with $HST$/WFPC2 images was collected from the $HST$ archive, and 
examined. 
Hereafter, we refer to this sample as the ``IRAS'' sample.   

Our goal is to provide a reliable assessment of the frequency of dust and dusty disks 
in early-type galaxies and how their morphology, amount, and dynamics relate to the activity and
other characteristics of the host galaxies. The main strength of our snapshot study is the 
large and unbiased sample of galaxies, all imaged at similar high resolution. This will allow us
to make reliable statistical statements regarding their dust properties.
Ultimately, we wish to use those with dusty nuclear
disks as a probe of the central BH masses, and study how they are related to
their environment. For consistency with Paper I, we use $H_o$ = 80 \hubu~throughout the paper.

\section{Analysis} \label{ana}

\subsection{Dust Morphology and Mass} \label{dustm}

Some general properties of galaxies in the snapshot sample are listed in Table \ref{gentab}. 
Paper I describes the observations and reduction of the images.
We now describe our analysis of the galaxies with dust. 
Paper I reports the detection of dust in 29 galaxies, or 43\% of the 
snapshot sample (the method of dust detection is outlined below).
This is generally comparable to those reported by previous studies of early-type
galaxies, which have shown the following detection rates of dust:
\citet{sg85}:$\sim$ 40\%;
\citet{vv88}: 23\%;
\citet{gou94}: 41\%;
\citet{dkfr95}: 48\%;
\citet{fer99}: $\sim$ 75\%;
\citet{tom00}: 56\%.
The large variation in the dust detection rates among different studies may be due to the
different methods of counting detections, or the different resolutions 
and sensitivities of the observations. 
The ground-based CCD study of \citet{vv88}, for example, includes only dust at the level of
$\gtrsim 10^4$ \msun, effectively missing all the small nuclear dust disks, resulting in an 
abnormally low detection rate. 
The relatively high detection rate of \citet{fer99}, on the other hand, may be due to higher 
sensitivity and lower detection threshold.
Also, we note that the 85\% dust detection rate found by \citet{vdb94} for galaxies in the Virgo 
cluster is considerably higher than those mentioned above. The reason for this is because they 
included galaxies with weakly distorted isophotes, in which the presence of dust was not directly 
visible by eye. In several cases, it later turned out that these distortions were not due to dust, 
but rather to defects in the CCD chips (so called ``measles''). Van den Bosch et al. (1994) 
detected dust ``by eye'' in 5 out of 14 galaxies (36\%), which is in good agreement with other studies. 

For our sample, the morphology of the dust is grouped into two broad categories:
filamentary and disky. Of these systems, 17 have only filamentary dust with no disks, 
9 have well-defined dusty disks, and 3 show dust lying not only in a disk-like
morphology but also in widespread extended filaments. 
See Table \ref{jdust} for a summary of the dust morphology found in the sample.
In Paper I, the level of filamentary dust is described in a purely qualitative way, 
with the simple scale I, II or III being assigned to denote the least to most dusty
galaxies through visual inspection only. Class I represents small traces of dust that do not 
greatly affect the isophotal shape of the galaxy; class II and III denote large amounts of
dust that prevent a meaningful analysis of the isophotes and luminosity profiles.
Figures \ref{ddisk} and \ref{fdust} show examples of some of the galaxies with dusty disks 
and filaments.
In this paper, we follow \citet{sg85} and \citet{dkfr95} to estimate the mass of the 
dust present in each system. The dust mass is given by:
\begin{equation}
M_d = \langle A_V\rangle\Sigma \Gamma^{-1} \label{mdopt}
\end{equation}
where $\langle A_V\rangle$ is the mean visual extinction in magnitudes, $\Sigma$ is the surface 
area covered by the dust, and $\Gamma$ is the extinction coefficient per unit mass.
We adopt $\Gamma = 6\times 10^{-6}$ mag kpc$^2$ \msun$^{-1}$ \citep{dkfr95}. 

\begin{figure*}
\caption{Examples of dusty nuclear disks. NGC 4125 and NGC 5813 contain both disky and 
filamentary dust. Compass arrow shows the direction of north. \label{ddisk}}
\end{figure*}

\begin{figure*}
\caption{Examples of galaxies containing filamentary dust of various levels. Both ESO 437$-$15
and ESO 580$-$26 are classified as F3; NGC 4589 is an F2 galaxy and NGC 3377 is F1, displaying
a single thin strain of dust in the upper left region.
\label{fdust}}
\end{figure*}

\begin{figure*}
\plotone{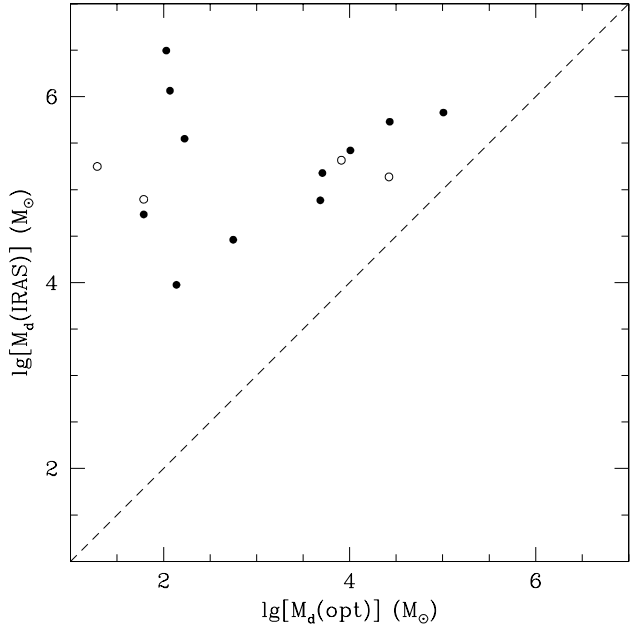}
\caption{Relationship between the optically derived dust mass, $M_d(opt)$, and the dust mass estimated from
$IRAS$ fluxes, $M_d(IRAS)$. Open circles denote dusty disks and filled circles denote filamentary dust.
The dashed line represents $M_d(IRAS) = M_d(opt)$. As can be seen, for galaxies with both visual dust
and $IRAS$ detections, $M_d(opt)$ is always lower than $M_d(IRAS)$ by 1--4 orders of magnitude. 
\label{mdoir}}
\end{figure*}

To obtain $\langle A_V\rangle$, we first determine the optical depth from the ratio of the
calibrated image to the model of its surface brightness generated from an 
isophotal elliptical fit, using the IRAF task ``ellipse'' with the dusty features masked out.
The optical depth in $R$ is then $\tau = - ln (F_{obs}/F_{mod})$, where 
$F_{obs}$ is the observed flux and $F_{mod}$ is the modeled flux,
and $A_R = 1.0857\tau$.
Assuming a Galactic extinction law with $R_V = 3.1$, we have $A_V = 1.33A_R$ (Cardelli, Clayton 
\& Mathis 1989). 
In determining the mean visual extinction, we include all areas where $\tau \gtrsim 0.02$, which
is the minimum detectable $\tau$.
Along with the estimated mass of the visible dust, Table \ref{jdust} lists the morphologies of 
the dust features, the mean visual absorption, the position angle (PA) of the major axis of the dust 
feature, and the PA of the galaxy major axis.
Table \ref{jdust} shows that the range of dust mass values for our sample is comparable to
that obtained by \citet{gdj95} and \citet{dkfr95}, typically 10$^1$--10$^5$ \msun\footnote{\citet{dkfr95}
refer to the total $M_{gas} + M_{dust}$ as the dust mass, with $M_{gas}/M_{dust} = 130$.}.
In Table \ref{mdcomp}, we show a comparison of the optically derived gas mass estimated by
our study with those of previous works. 
For consistency, we consider only results obtained from similar $HST$ observations by
\citet{dkfr95} and \citet{tom00}. A gas-to-dust ratio of 130 is assumed for our study, and 
previous mass values have been converted using distances listed in Table \ref{gentab}.
Although the number of objects in common between these studies are small, the mass estimates
are in reasonable good agreement, given the uncertainty of $\sim$ 0.5 in logarithmic 
scale \citep{tom00} typical for this type of dust mass determination. The agreement appears better
with \citet{dkfr95}, who used the same method as our study, than with \citet{tom00}, who used
color excess, giving mass values that are a few times to one order of magnitude larger than ours. 

Note that the dust mass estimated above relies on the simple assumption of a foreground
screen in front of a background light. Any dust embedded within the galaxies or 
intermixed with stars cannot be estimated by this method. For example, using radiative transfer 
calculations, including scattering of photons into the line of sight, \citet{mar00} have found that 
the mass estimated for the dusty disk in NGC 4261 is about one order of magnitude greater than that 
inferred from a foreground screen model. 
Furthermore, since we are fitting directly to the observed light, and it is impossible to mask out 
all faint dust features, we do not have a perfect model of the true underlying light.
Thus the derived masses listed in Table \ref{jdust} are always {\it lower} limits to the true mass. 

Some of the missing non-clumpy, diffuse dust could be detected through its far-infrared radiation.  
This is confirmed by our estimates of the dust mass from the $IRAS$ flux densities $M_d(IRAS)$. 
Following \citet{gdj95}, the dust mass is estimated from the $IRAS$ flux through the
formula:
\begin{equation}
M_d = 5.1\times 10^{-11} S_\nu D^2 \lambda^4_\mu(e^{1.44\times 10^4/\lambda_\mu T_d} - 1)~M_{\sun}
\label{mdiras}
\end{equation}
where $S_\nu$ is the $IRAS$ flux density in mJy, $D$ is distance in Mpc, and $\lambda_\mu$ is in
$\mu$m. The dust temperature $T_d$ is estimated from the color ratio $S_{60}/S_{100}$, according
to the prescription of \citet{kx92}, using an emissivity law that varies as $\lambda^{-1}$. 
For those galaxies where the color ratio $S_{60}/S_{100}$ is not available, a representative color
temperature of $T_d = 30$ K is assumed. 
We caution that the $IRAS$ dust mass is very sensitive to such assumed parameters as grain size and 
temperature.
Table \ref{jrir} lists the radio and $IRAS$ fluxes for those galaxies detected in the sample. The radio
1400 MHz flux densities are derived from the NRAO VLA Sky Survey (NVSS; Condon et al. 1998), and the 
$IRAS$ fluxes are taken from \citet{knap89}, except for ESO 437-15, ESO 447-30, ESO 580-26, and NGC 2824,
which are not included in their sample. For these galaxies, we measured their flux densities from the 
data scans by using the {\bf apphot} package in IRAF and comparing the results to values in the 
$IRAS~Faint~Source~Catalog$ to derive scaling factors, which are similar to those of \citet{knap89}.
As Figure \ref{mdoir} shows, for those galaxies with both $IRAS$ detections and visibly detectable dust, 
the dust masses estimated from the $IRAS$ fluxes, 
listed in Table \ref{jrir}, are $\sim$ 1--4 orders of magnitude greater than those 
derived from optical depth measurements of visual dust structures.
This is consistent with earlier studies \citep{gdj95,dkof00}.

\subsection{Correlation with Galaxy Properties} \label{galcorr}

\begin{figure*}
\plotone{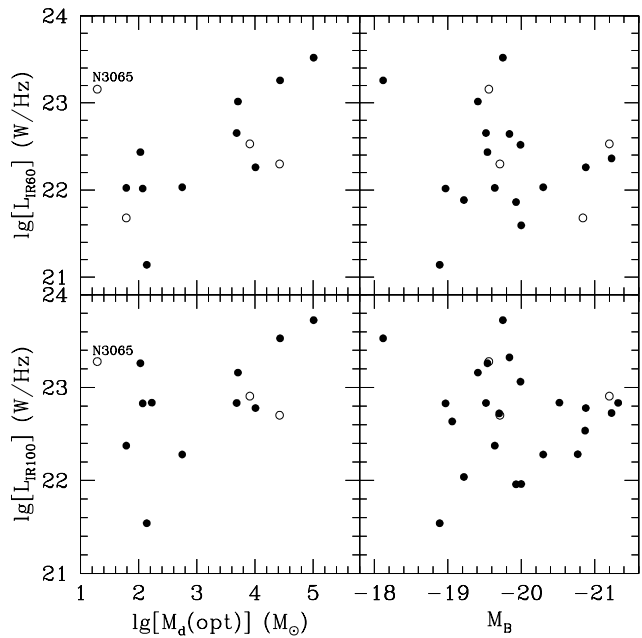}
\caption{$IRAS$ 60$\mu$m and 100$\mu$m luminosities versus absolute blue magnitude $M_B$ and dust mass 
as determined from optical depth measurement $M_d$(opt). Symbols are as in Figure \ref{mdoir}. There 
appears to be a significant correlation between $IRAS$ luminosities and dust mass, especially 
when only filamentary dust systems are considered, but no correlation between $IRAS$ luminosities 
and $M_B$. The only outliner is NGC 3065, whose $IRAS$ fluxes are likely contaminated with emission 
from a nearby source. \label{irall}}
\end{figure*}

\begin{figure*}
\plotone{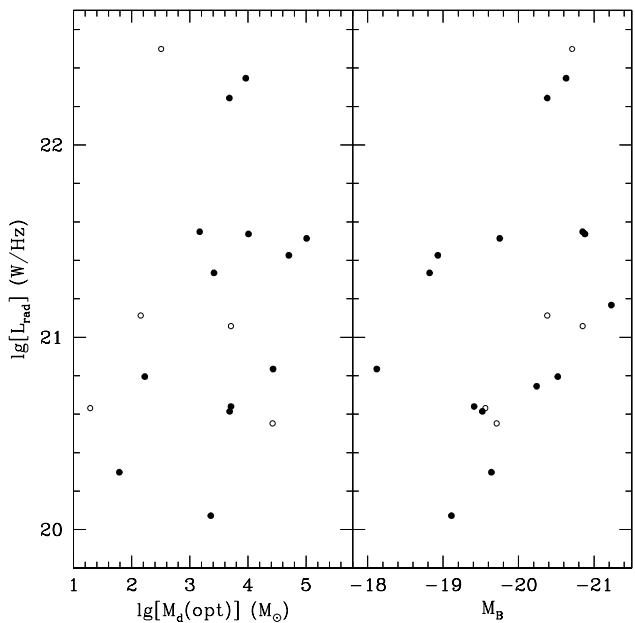}
\caption{Radio luminosities at 1.4 GHz as detected by NVSS, $L_{rad}$ versus dust mass and absolute blue
magnitude. Symbols are as in Figure \ref{mdoir}. There is a weak correlation between $L_{rad}$ and 
$M_d$(opt), and a significant anticorrelation between $L_{rad}$ and $M_B$. \label{nvss}}
\end{figure*}

Although the optically derived dust mass, $M_d(opt)$, always underestimates the true dust mass, we 
choose to use it in our analysis, since the $IRAS$ derived dust mass, $M_d(IRAS)$ is unsuitable for any 
analysis relating to the far-IR properties. If derived in a consistent manner for the whole sample, 
$M_d(opt)$ should be a reliable representation of the dust mass. Moreover, some galaxies have no $IRAS$ 
detections and yet dust is detected visually (see below). Presumably, most of the dust in these objects 
may be too cold ($\lesssim$ 20 K) to be detected by $IRAS$ \citep{tm96,mer98}.
Figures \ref{irall} and \ref{nvss} show the scatter plots of various properties of the snapshot sample
galaxies versus dust mass.
Table \ref{rtab} summarizes the correlation coefficients for these plots. 
As can be seen, both the 60$\mu$m and 100$\mu$m luminosities are significantly correlated with dust 
mass. Furthermore, the correlation becomes stronger when the few systems with dusty disks are excluded. 
The only outliner is NGC 3065 whose IR fluxes are suspected to be contaminated 
with emission from NGC 3066, 3\arcmin~away \citep{knap89}. This source is excluded from our 
statistical analysis.
As Figure \ref{irall} and Table \ref{rtab} also show, there appears to be no 
significant correlation between $IRAS$ luminosities and absolute blue magnitudes of these galaxies. 
Turning to the radio property, Figure \ref{nvss} and Table \ref{rtab} show that the 1.4 GHz radio 
luminosity is also seen to follow a weak, but significant relationship with dust mass.
However, in this case, the radio luminosity is also correlated with the optical luminosity of 
the galaxies, in agreement with Calvani, Fasano, \& Franceschini (1989).
This suggests that the $log L_{rad}$ -- $log M_d$ correlation may be largely a secondary
effect of a more fundamental relationship between optical luminosity (and thus mass) and radio 
luminosity of the galaxy.

\begin{figure*}
\plotone{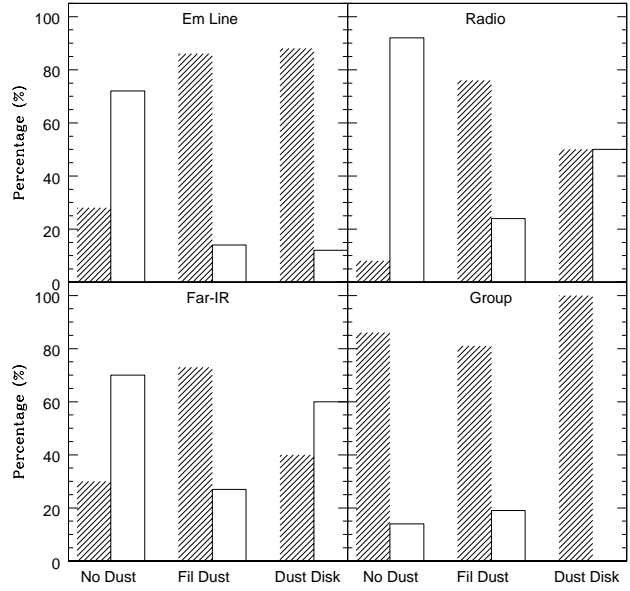}
\caption{Detection rates of emission lines, radio 1.4 GHz emission, $IRAS$ 60 $\mu$m and 100 $\mu$m 
emission, and being in group/cluster for three groups of galaxies in the snapshot sample: 
those with no dust, filamentary dust, and dusty disks. Shaded bar denotes detection, and empty bar 
denotes non-detection. Percentages are with respect to the number of galaxies in each group with 
available data. A much greater fraction of dusty galaxies has emission line, radio and $IRAS$ detections 
than dustless galaxies. There is essentially no difference in the preference of being in group/cluster 
for galaxies with or without dust. \label{pdet}}
\end{figure*}

If one considers the radio detection rate, not constrained solely to those galaxies found to have 
dust, the correspondence of radio-emitting sources to the presence of dust is much stronger. 
Figure \ref{pdet} and Table \ref{drate} present the detection rates of various properties for both
the snapshot and $IRAS$ samples.
In the snapshot sample, only 8\% of the galaxies without dust have radio detection, while 66\% of the 
galaxies with dust do. 
Similarly, of the 26 galaxies in the snapshot sample with $IRAS$ 100 $\mu$m 
detection, most (58\%) are associated with dusty galaxies, a great majority of which are filamentary 
(Table \ref{drate}).
Since the galaxies are selected to be nearby ($\lesssim$ 40 Mpc), any distance effects are likely to 
be small. The overall rate of $IRAS$ detection in the snapshot sample is 42\%, consistent with those 
reported by \citet{knap89} for E and S0 galaxies. 

Turning to the $IRAS$ sample listed in Table \ref{irastab}, about 3/4 (78\%) of the galaxies contain dust 
(i.e., detection by eye from $HST$ images) of some type, compared to only 43\% for the snapshot sample. 
Since the $IRAS$ sample is specifically selected for the detection of 100 $\mu$m emission, this suggests
that the far-IR radiation largely arises from dust embedded throughout the galaxy. 

\begin{figure*}
\plotone{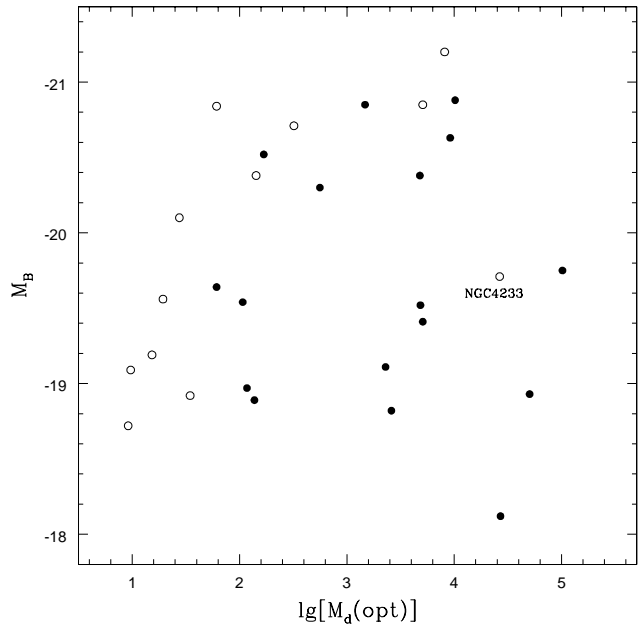}
\caption{Absolute blue magnitude $M_B$ versus optically derived dust mass $M_d$(opt). Symbols are as 
in Figure \ref{mdoir}. No relationship exists when all systems are considered, but a significant 
correlation is evident for the dusty disk systems. \label{dustmb}}
\end{figure*}

As Figure \ref{dustmb} shows, there appears to be no correlation between the derived dust mass and 
absolute blue magnitude. This result is consistent with that of \citet{gdj95} and 
\citet{dkfr95} but disagrees with \citet{mer98} and \citet{fer99}. 
At first sight, this seems to suggest that there is little causal 
connection between the dust and the existing stellar population in the galaxy.
However, if only the 12 galaxies with dusty disks are examined (open circles in Fig. \ref{dustmb}), 
a clear relationship 
is evident between the dust mass and blue luminosity of the galaxies. The correlation between
$M_d$ and $M_B$ is highly significant, with a Spearman rank coefficient of $r_s = -0.76$ ($p< 0.007$)\footnote{Since the
correlation does not appear linear, we quote the Spearman rank coefficient here; a Pearson linear 
coefficient is given in Table \ref{rtab} for comparison.}. 
NGC 4233 is an outliner, reducing the the significance of the correlation. Excluding NGC 4233 
improves the correlation coefficient to $r_s = -0.88$ ($p < 0.0008$). 

\subsection{Correlation with Velocity Dispersion} \label{vdis}

\begin{figure*}
\plotone{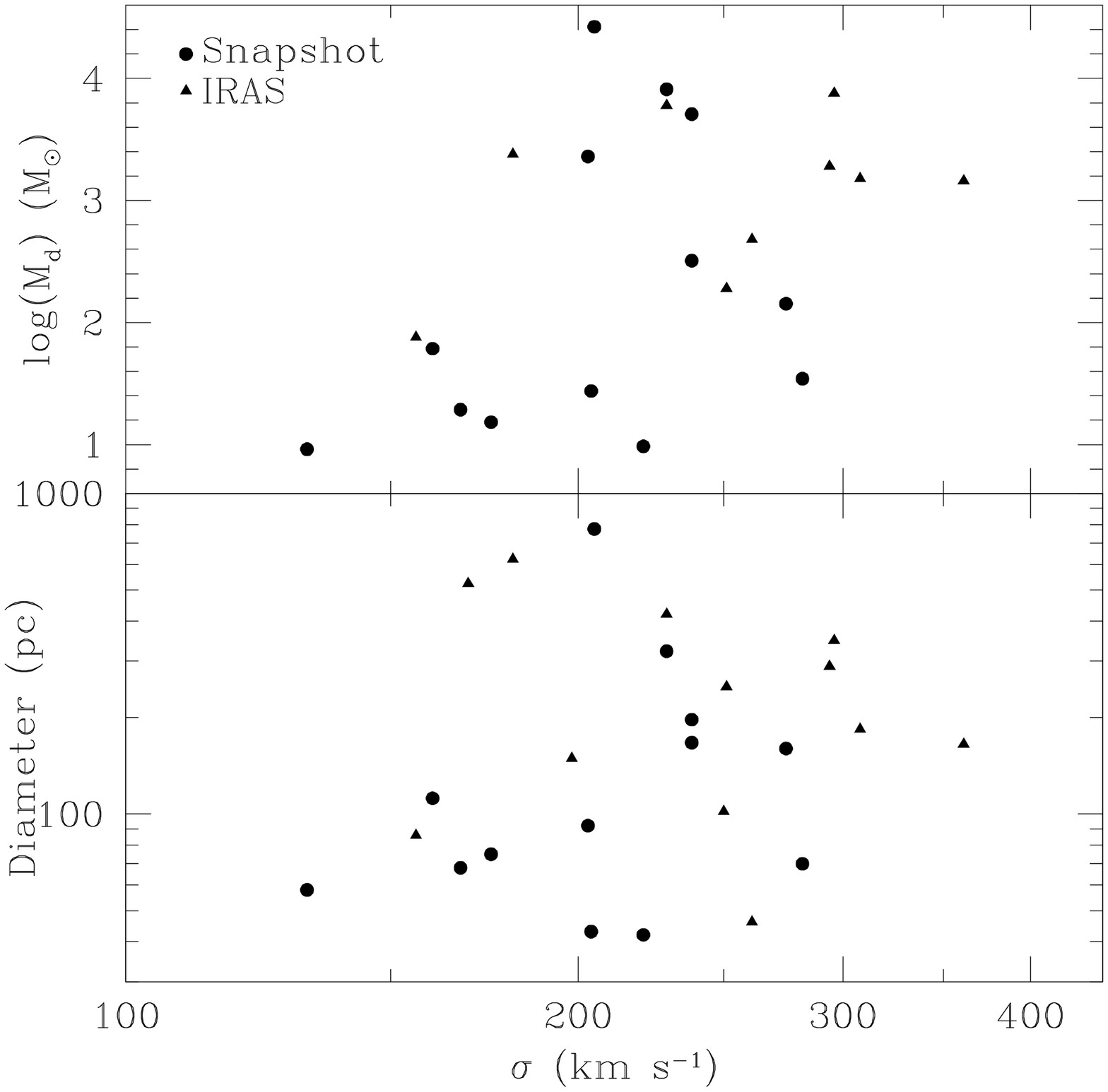}
\caption{Relationship between the mass ({\it top}) or diameter, $d$ ({\it bottom}) of the dusty disks
and central velocity dispersion of the galaxy, $\sigma$. Filled circles denote galaxies from the snapshot 
sample, and filled triangles represent those from the $IRAS$ sample. A weak but significant correlation is 
seen, especially for those systems with $d \lesssim 400$ pc. \label{mdsizesig}}
\end{figure*}

The mass of the central BH in early type galaxies has been shown to be correlated with
the bulge luminosity (and hence its mass;
e.g., Kormendy \& Richstone 1995; van der Marel 1999). Very recently, a much tighter correspondence 
has been shown between the BH mass and the velocity dispersion of the galaxy 
\citep{fm00,g00}. 
In Figure \ref{mdsizesig} we show plots of the mass and diameter of the dusty disks versus the central 
velocity dispersion $\sigma$ for both galaxies in the snapshot and $IRAS$ samples. $\sigma$ is the
velocity dispersion taken from \citet{dav87} and \citet{mcel95}, and we have used 
distances derived from surface brightness fluctuation (SBF; e.g., Neilsen \& Tsvetanov 2000) wherever 
possible. There appears to be a weak tendency for larger disks with higher dust mass to show
higher velocity dispersion. However, the diagrams contain considerable scatters, and any correlation 
is very weak at best. Notable are the apparent ``zones of avoidance'' in the upper left and lower
right areas of the graphs. 
A formal test shows that the $M_d$ vs. $\sigma$ plot has a correlation 
coefficient of $r= 0.45$ ($p[null] = 0.035$). 
Interestingly, in the disk size vs. $\sigma$ diagram, the correlation seems to apply only for 
those galaxies whose dust disk diameter is $\lesssim$ 400 pc. 
Those galaxies with very extended ``disks'', or more like lanes, such as
NGC 5128 (Cen A), NGC 4697, NGC 4233, NGC 6861 and ESO 208-21 lie well above the trend
line. Discarding these from the analysis, statistical test shows that the correlation 
has a coefficient of $r = 0.49$ ($p[null] = 0.022$). 

If the correlation is real, it is interesting to speculate on the cause and implications of such a 
relationship. The more luminous galaxies with larger mass may be expected to contain more dust. 
Indeed, there is a strong relationship between dust mass and absolute magnitude
for those galaxies with dusty disks (Fig. \ref{dustmb}). 
The loose correlations between the disk size/mass and $\sigma$ 
(Fig. \ref{mdsizesig}) could be a direct consequence of a correlation between galaxy mass and $\sigma$, 
which has long been documented \citep{fj76,f87}. 
The remarkable sharpness of the disks detected (Fig. \ref{ddisk}) is also highly 
suggestive of some physical mechanism that acts to limit the size and shape of these disks. If so, the 
disk size/mass -- $\sigma$ relation indicates that such process is ultimately governed by the mass 
of the central BH. 

\subsection{Origin of the Dust} \label{dori}

\begin{figure*}
\plotone{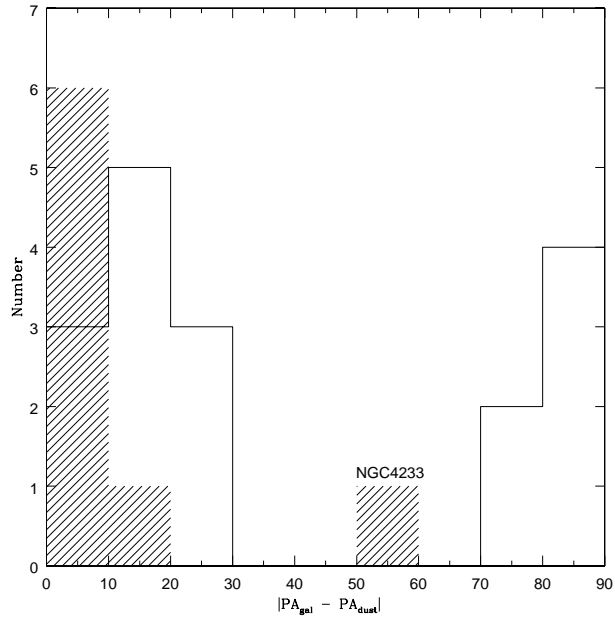}
\caption{The distribution of the difference between the PA of the major axis of the galaxy and 
the PA of the major dust structure. Shaded histogram denotes dusty disks and empty histogram 
denotes filamentary dust systems. Systems with low inclinations (NGC 3065, NGC 4648, NGC 5017, NGC 5812) 
are excluded. Dusty disks are much more well aligned with the host galaxy than filamentary dust. 
\label{padiff}}
\end{figure*}

Clues to the origin of the dust can be glimpsed from a comparison of the PA between the major 
axis of the dust features and that of the galaxy.
The isophotal PA of the galaxy's major axis is measured at 10\arcsec, and listed in Table \ref{jdust}.
Table \ref{jdust} also lists the PA difference $\vert \Delta PA\vert = \vert PA_{g} - PA_{d}\vert$ 
between the two major axes. Figure \ref{padiff} displays a distribution of the PA difference.
As can be seen, the dusty nuclear disks have a high tendency to be aligned with the
major axis of the host galaxies, nearly all within 10\arcdeg, while the filamentary dust 
structures show a much lower tendency to align with their hosts. The former result is
in agreement with that of \citet{mar00} for a sample of dusty disks in nearby 3CR elliptical galaxies.
The median $|\Delta PA|$ for 
the dusty disks and filamentary dust are 5\arcdeg~and 23\arcdeg, respectively.
A K-S test confirms that the $|\Delta PA|$  distributions are significantly different, with
only a 0.7\% probability that they are drawn from the same parent population.
This may suggest different origins for the disky and filamentary dust: internal for the dusty disk
and external for filamentary dust. The presence of a correlation in Figure \ref{dustmb}
for dusty disks but not for filamentary dust also seems to corroborate this.
However, if both types of dust have a common origin, the most likely interpretation is that 
the dust comes from the outside and requires time to settle into well-organized disks.  
Again, the outliner is NGC 4233, which is the only galaxy with dusty ``disk''
that does show a misalignment ($\Delta PA = 52\arcdeg$). This galaxy, however, 
happens to have very extended double dust lanes, rather than the more typical 
compact nuclear disk, perhaps suggesting that the dust has been acquired more recently and
is still in the process of settling into a compact nuclear disk. 

For those galaxies that have filamentary dust, the acquired dust
has not yet had sufficient time to relax and achieve equilibrium in a ``settled state''. 
If the dust had arisen internally we would have expected a general alignment with
the major dynamical axis of the galaxy.
Our finding is consistent with that of \citet{dkfr95}, who found a  
kinematic misalignment between the stars and the dust in a sample of 
early-type galaxies, suggesting that the two are not kinematically coupled, 
and thus have different origins.
In a study of a sample of radio-loud active galaxies (the 3CR sample), \citet{dkof00} also
find the dust in various forms similar to those seen here, and they suggest that these are also 
probably reflective of their different dynamical states. 
Previous studies have reached similar conclusion that the dust is external in origin
\citep{k89,f91,gdj95}.

\section{Discussion} \label{diss}
 
\begin{figure*}
\plotone{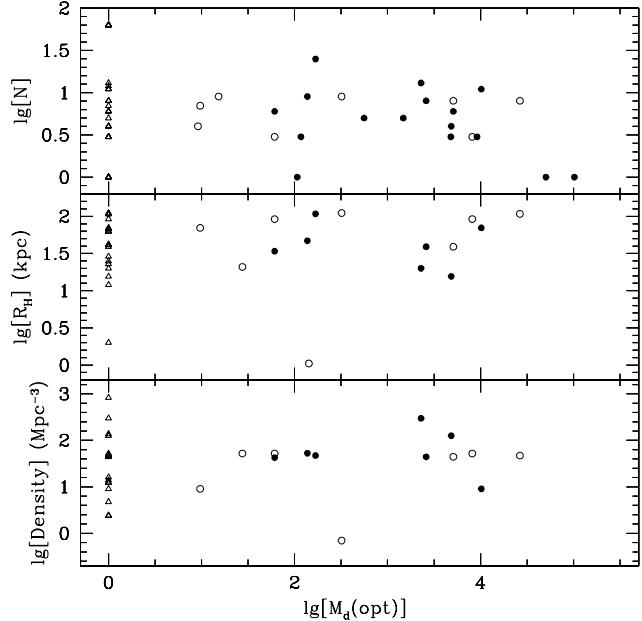}
\caption{The number of group members $N$, mean harmonic radius $R_H$, and number density of 
the group as a function of dust mass for the snapshot sample. Symbols are as in Figure \ref{mdoir}. 
In addition, open triangles denote galaxies with no detectable dust. No correlation is evident between 
these quantities. \label{group}}
\end{figure*}

If the dust originated in a merger of two gas-rich disk galaxies (Barnes \& Hernquist 1996, 1998), 
one might expect that the stars have not fully reached equilibrium either. However, no 
evidence for recent merger activity has been found in stars, even in the outer parts of the galaxy. 
This may suggest that the dust could arise from the tidal capture of gas from an encounter
with a nearby galaxy. 
As a simple test of this hypothesis we have looked for any correlation between the
amount of dust and local environment density of the galaxies in the snapshot sample. 
The group membership is taken from \citet{gar93}, \citet{hg82} and \citet{gh83} and 
are listed in Table \ref{gentab}.
As a rough measure of the size of the group we have also considered the 
group harmonic radius taken from \citet{hg82} and \citet{gh83}. It is defined as
$R_H = \frac{\pi V}{H_0} sin \{ [N(N-1) \sum_{j<1} \sum_{i=1} \frac{1}{\theta_{ij}} ]^{-1} \}$, 
where $V$ is the mean velocity of the group, $N$ is the number of group members, and $\theta_{ij}$
is the angular separation of the $i$th and $j$th group members.
We show in Figure \ref{group} the scatter plots of $M_{d}$ versus $N$, $R_H$ , and the 
number density of the group defined as $N/\frac{4}{3}\pi R_H^3$. As can be seen, there appears
to be no correlation between the dust mass with any of these quantities.

As Table \ref{drate} and Figure \ref{pdet} also show, there is virtually no difference between 
the extra-galactic environment of the two types of galaxies: dusty galaxies are no more likely to be 
found in groups or having close companions than dustless galaxies. 
These tests, therefore, fail to support an external origin for the dust.
However, given other evidence for an external origin from this and
previous studies, it may be possible for the dust to either be captured sufficiently long 
ago that no signs of mergers or interactions are currently evident, or be acquired more recently 
but has an efficient way of losing angular momentum quickly.
As shown by Tohline, Simonson, \& Caldwell (1982) and \citet{scd88}, the timescale that dissipative 
differential precession causes gas entering an elliptical galaxy to settle in a disk is fairly short 
($\sim 10^9$ yrs). 
We note that an external origin, coupled with the $|\Delta PA|$ distribution in Figure \ref{padiff}, 
also suggests that elliptical galaxies are predominantly oblate \citep{tsc82}.
The external or internal origin of the dust could be confirmed with a systematic spectroscopic study 
of a large and well-defined sample of galaxies with dust, such as that of the present study, to 
examine the dynamics and sense of rotation of the stars and gas. If approximately half of the disks 
are counter-rotating with respect to the stars, the dust most likely has an external origin.
Such counter-rotating disks have been observed in a number of galaxies \citep{bc99,c00}, 
with a higher incidence in early-type galaxies \citep{kf00}, providing support for the
idea that these disks are products of galaxy interactions or mergers.
 
It has been shown that the presence of dust is closely associated with ionized gas in early-type
galaxies \citep{c84,gou94,fer99}. 
The presence of ionized gas alone, however, does not establish the object as being a {\it bona fide} 
active galactic nuclei (AGN), with the excitation of the gas provided by photoionization 
from a central nonthermal continuum. Because active star forming regions are often associated with dust, 
in some of these systems, the gas ionization probably comes from circumnuclear starburst or 
high-velocity shocks and not a true active nucleus (i.e, gas-fed BH), although many are found 
to be LINERs (Table \ref{gentab}; many of which are believed to be low-luminosity AGNs).
For the purpose of the following discussion, we shall take the 
presence of any emission line as a sign of galaxy activity. 

We wish to see if ionized gas indicative of galactic activity is more reflective of not only dust, 
but dust in a small nuclear disk.
That is, is it more likely for a galaxy to show emission lines when a dusty disk is present
than when dust is filamentary in form or no dust is seen?
It is natural to expect galaxies with dusty nuclear disks to show
some level of nuclear activity because there is an ample gas supply from the disk to feed
the central massive BH, if it is present.
Furthermore, the closer the dust is to the nucleus, the more likely the connection is to the activity. 
To answer this question, we have searched the literature 
for observations of any line emission in the snapshot sample.
Table \ref{gentab} summarizes our results. 
Of the 67 galaxies in our snapshot sample, we found 43 with published observations establishing whether or
not ionized gas is present. 
Our search indicates that 19 (86\%) of the dusty galaxies do show
line-emitting gas, compared to only 6 (28\%) of the dustless galaxies. 
As \S \ref{dustm} shows, however, if no dust is visually detected, it may not mean that it is truly 
free of dust; the dust could be in 
a diffuse, non-clumpy form that is not visibly apparent. 
Figure \ref{pdet} indicates that the presence of line-emitting gas does not show any preference for 
galaxies with dust in the form of a small disk, or extended clumps/filaments. In fact, the incidence 
of emission lines is virtually identical ($\sim$ 87\%) between the two groups. 
This indicates that emission line activity does show a strong preference for galaxies with {\it visibly 
detectable} dust, consistent with the finding by \citet{c84} and \citet{tom00}, but not necessarily
in small nuclear disky form.
Taken as a whole, emission lines occur in about 58\% of the snapshot sample galaxies. 
This is consistent with \citet{ho97a}, who found that the detection rate for emission line in E and S0 
galaxies in a large, complete sample of nearby galaxies is 60\%. 

While most of the dusty disk galaxies in our sample have ionized gas, we have found 
at least a few without any signs of line emission (e.g., NGC 4406, NGC 4648). While this may indicate 
that the emission level is below the detection limit of the surveys, how the fuel is being fed to the 
central BH and the rate of accretion may be an important factor in determining the energy production
and emission. Perhaps, a central massive BH may not have yet formed in some galaxies
even though a dusty disk is present; alternatively, the BH is present but the accretion rate may not 
be high enough to ignite the activity: the absence of activity may be due to either a puny BH 
or an underfed massive one. For the two galaxies mentioned above, their velocity dispersions of order 
240 \kms~\citep{mcel95} suggest a black hole mass of $M _{BH} \approx 2 \times 10^8$ \msun, quite 
typical of many early-type galaxies \citep{fm00,g00}. Thus, their inactivity may most likely be due to
a radiation inefficient mode of, or lack of, fuel accretion rather than a lack of a BH.

Further support for a lack of fuel transport comes from close examination of the central regions of 
dusty galaxies. \cite{mp99} examined WFPC2 images of the CfA sample of Seyfert 2 galaxies
and found that the presence of nuclear spiral dust lanes within the inner 10-100 pc is nearly
ubiquitous. Apparently, such dust spirals appear necessary in order for the gas to lose angular
momentum and infall effectively into the supermassive BH, thereby producing real AGN activity. 
Although we were unable to produce color images for most of our galaxies since they were observed 
in only one band (F702W), examination of the optical depth images (see \S \ref{dustm}) revealed that 
nuclear dust spirals are not a common feature in our sample galaxies, appearing in only about four of 
29 dusty galaxies. This is consistent with their lack of strong activity. 

We showed in \S \ref{galcorr} that a significantly higher fraction of the galaxies with dust have 
detections in the radio. 
A survey of radio-loud early-type galaxies \citep{vk99} has shown that the 
incidence of dust is 89\%, about twice as high as found in our snapshot sample of relatively benign, 
radio-quiet early-type galaxies (43\%). 
Even in the $IRAS$ sample, which is biased for the detection of dust,
the dust detection rate (78\%) is significantly lower than that of the radio-loud sample. 
However, the overall rates of radio and emission line detection are slightly higher than
those in the snapshot sample (Table \ref{drate}).
Recently, \citet{s00} show that the detection rate of dust and dusty disks in nearby FR-I radio 
sources is nearly 100\%. In 4 out of 5 cases where the galaxies are sufficiently nearby to allow a 
detailed look of the nuclei, Sparks et al. also report the presence of optical jets emanating 
perpendicular to face-on dust disks. 
These studies provide strong support for the idea that radio and nuclear activities are intimately 
linked with the presence of clumpy dust, and strongly suggest that the disks and filaments are the 
source of fuel and gas supply for the radio engines lying at the core of the galaxy.
If all galaxies have central BHs (as is very likely) and the presence of dust is higher
in galaxies that are active, this implies that galaxies have active duty cycles, and that the
central supply of gas and dust is the requisite for the onset of the activity.

The probable relationship between the mass/size of the dusty disks and the central velocity 
dispersion (\S \ref{vdis}), coupled with a similar correspondence between the latter
and the central BH mass, suggests that the mass of the disk is connected with that of the BH.
If real, it would imply that activity in early-type galaxies may undergo a process of evolution
in which dust is acquired, settles, and provides a channel and fuel
reservoir that support the growth and activity of the central massive BH. 
We note, however, that not all galaxies in which a central BH has been found show visibly 
detectable dusty disk, and not all galaxies with dusty disk are detected, due to projection 
effects \citep{dkfr95}. Thus, the two effects might mitigate each other to some extent. 

\section{Summary and Conclusions} \label{concl}

Confirming previous studies, we find that dust is very common in elliptical studies.
Although this result has been noted before, our unbiased and 
volume-limited sample puts such finding on a better footing. There is a strong 
correspondence between the presence of dust and the detection of radio and far infrared
emission. Among those galaxies with visibly detectable 
dust, a significant correlation is found between the derived mass of the dust and 
infrared luminosities, especially when only systems with filamentary dust are considered.
At the limit of the $IRAS$ survey, dusty galaxies are more than 10 times more likely to have 
100 $\mu$m detection than dustless galaxies, and those with filamentary dust are 
$\sim$ 4 times more likely to have 100 $\mu$m detection than those with dusty disks.
This result suggests that the 100 $\mu$m radiation most likely comes from filamentary 
dust which absorbs UV and visible radiation from the surrounding stars and reradiate
it in the far-IR. While it is possible to have detection in 100 $\mu$m {\it without} visually 
detectable dust (e.g., NGC 2549, NGC 3348), the presence of dust makes the detection much more likely. 

Galaxies with detectable, clumpy dust are also much more likely to show nuclear activity in 
the form of radio and line emission than dustless galaxies. 
The formation of clumpy dust thus appears to be a necessary ingredient for radio and emission-line 
activity. However, galaxies with dust in small disky morphology are not any more likely to show such
activity than those having irregular, filamentary dust.  
Nearly all dusty disks are well aligned with the major axes of the galaxies in which they
reside, while dust filaments seem to be more randomly oriented. If both types of dust have the 
same origin, then the dust is most likely acquired externally and requires time to settle dynamically. 
However, there appears to be no correlation between the presence of dust and the ambient galaxy density.
Alternatively, the two types of dust may have different origins. 
A spectroscopic survey to study the dynamics of a well-defined sample of galaxies is needed to confirm
the external or internal origin of dust. 
The possible relationship between the size of the dusty disks with the central velocity dispersion of
the galaxy suggests a causal connection between the BH mass and the mass/size of the dusty disk and bulge. 

\acknowledgments

This research is supported by NASA grants 6357 and 8386 from the Space Telescope Science 
Institute, which is operated by AURA, Inc., under NASA contract NAS5-26555. We thank the referee
for a number of very helpful suggestions. We have made use of the 
CATS database \citep{v97} of the Special Astrophysical Observatory, the
LEDA database (http://leda.univ-lyon1.fr), and the NASA/IPAC Extragalactic Database (NED), which is 
operated by the Jet Propulsion Laboratory, California Institute of Technology, under contract 
with NASA. 


\clearpage
\begin{figure}
\psfig{file=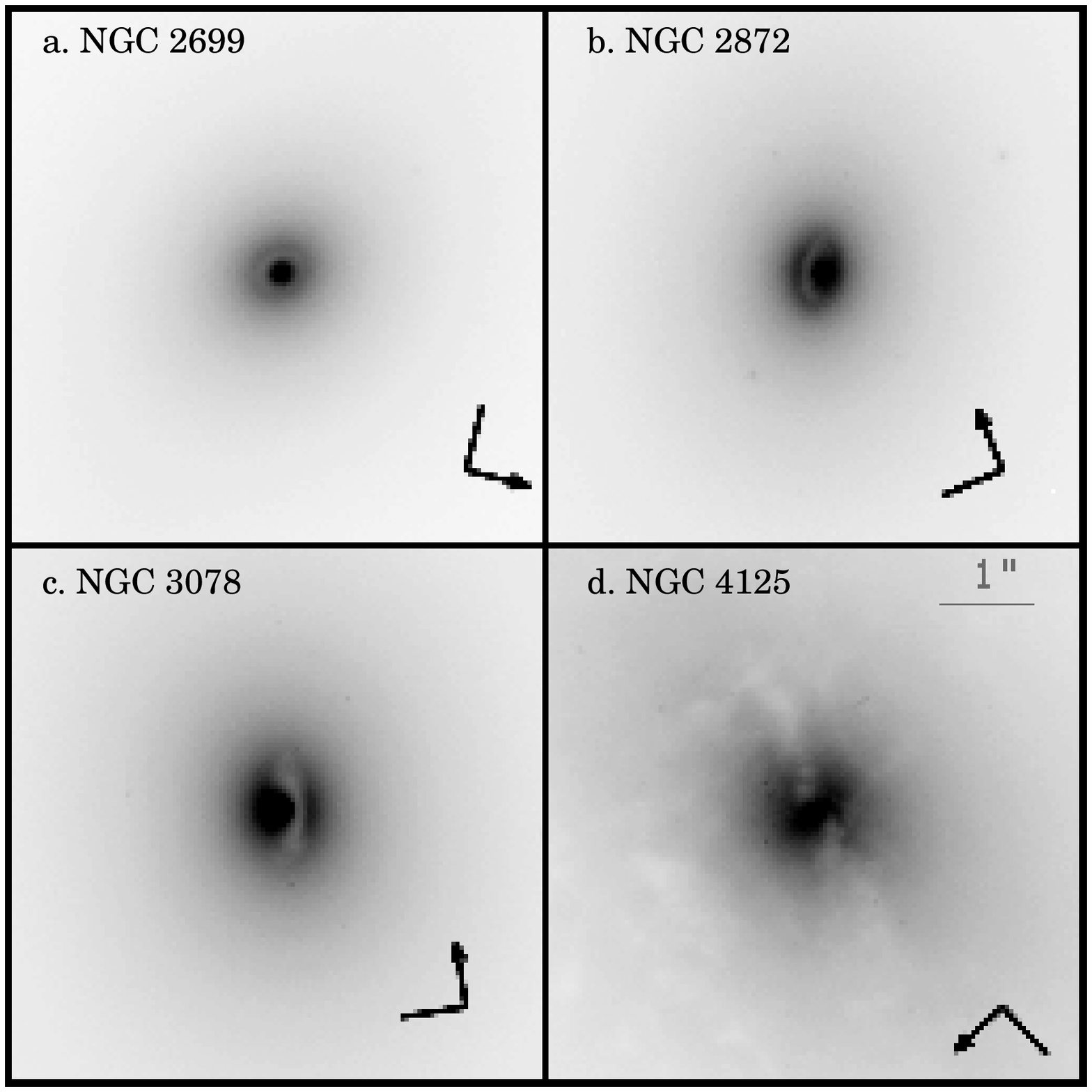,width=4.5in,height=4.5in}
\psfig{file=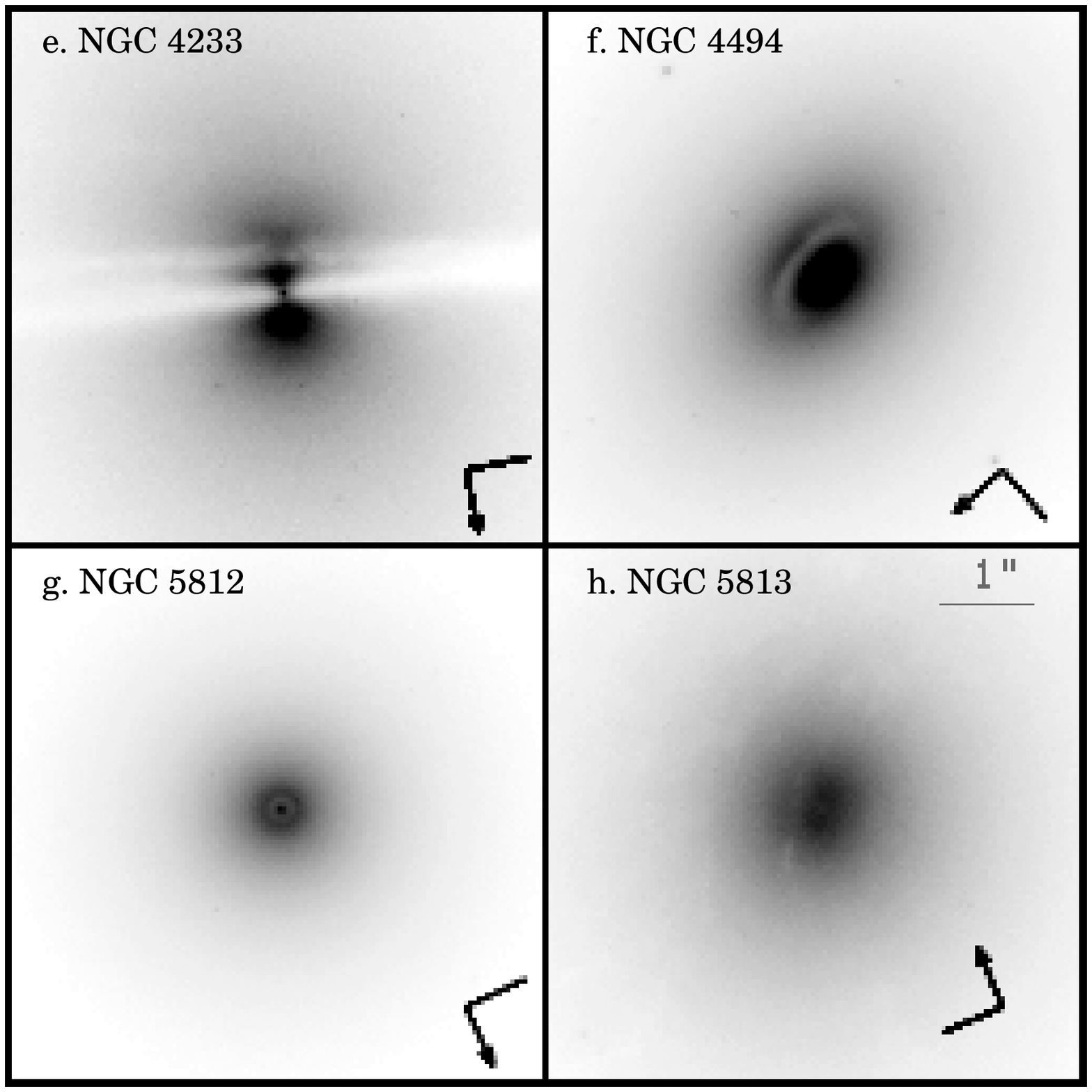,width=4.5in,height=4.5in}
\figurenum{1}
\caption{}
\end{figure}

\begin{figure}
\psfig{file=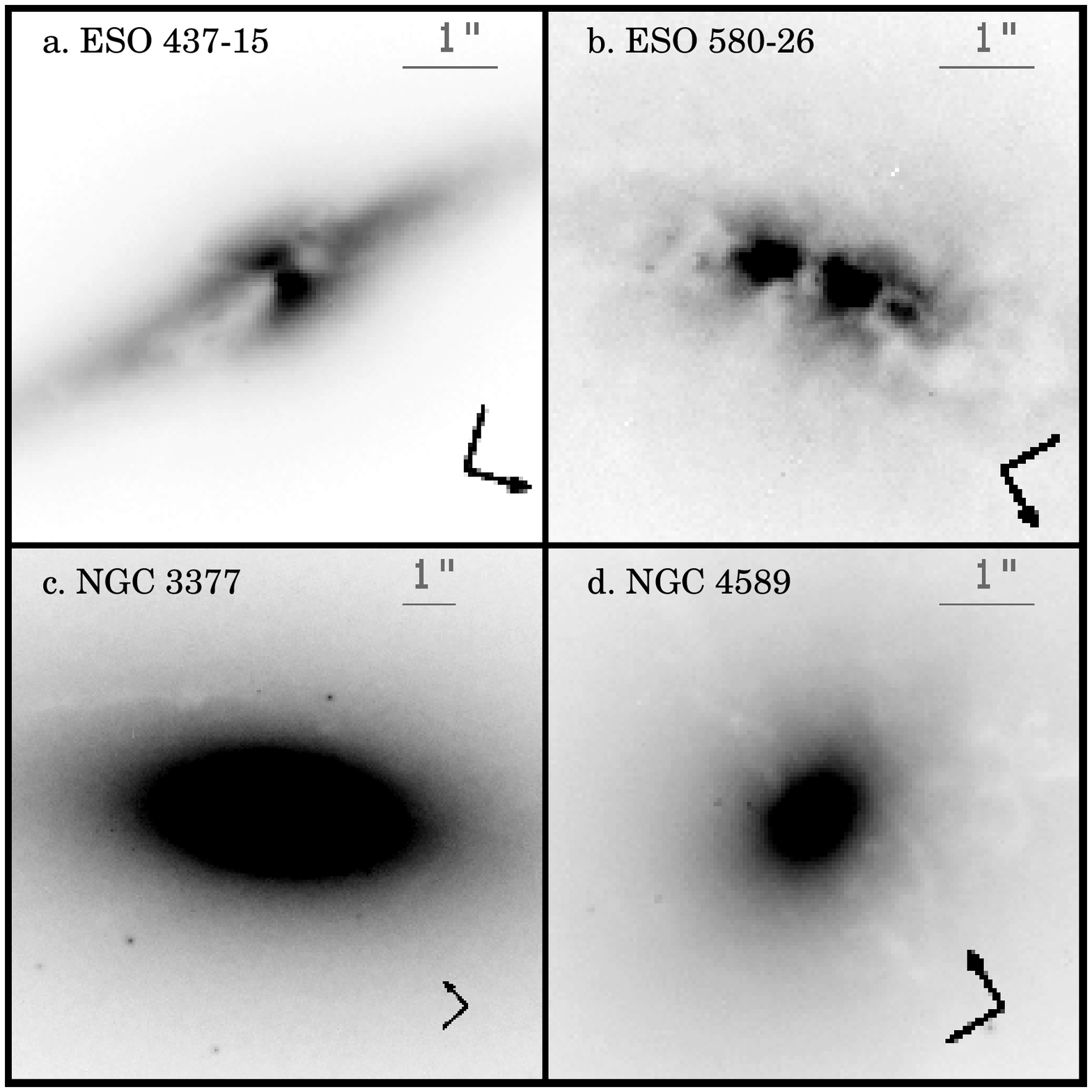,width=6.0in,height=6.0in}
\figurenum{2}
\caption{}
\end{figure}

\clearpage
\begin{deluxetable}{lccccccccccl}
\tabletypesize{\scriptsize}
\tablecaption{Snapshot Sample: General Properties \label{gentab}}
\tablewidth{0pt}
\tablehead{
\colhead{Name} & \colhead{$M_B$} & \colhead{$B_T$} & \colhead{$D$} & \colhead{Dust} & \colhead{Line Emiss} &
\colhead{LGG Group} & \colhead{N} & \colhead{CfA Group} & \colhead{N} & \colhead{$R_{H}$} & \colhead{Ref} \\
\colhead{(1)} & \colhead{(2)} & \colhead{(3)} & \colhead{(4)} & \colhead{(5)} & \colhead{(6)} &
\colhead{(7)} & \colhead{(8)} & \colhead{(9)} & \colhead{(10)} & \colhead{(11)} & \colhead{(12)} \\
}
\startdata
ESO 378-20    & $-19.58$ & $13.17$ & $ 35.6$ &   0   &   & LGG256 &  7  & ...    & ...  &  ...       & \\   
ESO 437-15    & $-19.41$ & $13.14$ & $ 32.3$ &   3   & N & LGG210 &  6  & ...    & ...  &  ...      & BM98 \\
ESO 443-39    & $-19.41$ & $13.39$ & $ 36.3$ &   0   &   & LGG328 &  6  & ...    & ...  &  ...       & \\ 
ESO 447-30    & $-19.84$ & $12.88$ & $ 35.0$ &   0   &   & ...    & ..  & ...    & ...  &  ...       & \\ 
ESO 507-27    & $-19.37$ & $13.56$ & $ 38.5$ &   0   &   & LGG310 & 12  & ...    & ...  &  ...       & \\   
ESO 580-26    & $-18.12$ & $14.85$ & $ 39.3$ &   3   & L & ...    & ..  & ...    & ...  &  ...      & OD85 \\ 
IC 875        & $-19.17$ & $13.73$ & $ 37.9$ &   0   &   & ...    & ..  & ...    &  2   &  ...       & \\   
MCG 11-14-25A & $-17.90$ & $15.32$ & $ 44.2$ &   0   &   & LGG238 &  3  & GH82   &  6   &  0.67      & \\   
MCG 8-27-18   & $-18.85$ & $14.36$ & $ 43.7$ &   0   &   & ...    & ..  & ...    & ...  &  ...       & \\   
NGC 2549      & $-19.22$ & $11.76$ & $ 15.7$ &   0   & N & ...    &  1  & ...    &  1   &   0       & H97 \\   
NGC 2592      & $-18.92$ & $13.11$ & $ 25.5$ &   4   &   & ...    & ..  & ...    &  3   &  ...       & \\   
NGC 2634      & $-19.99$ & $12.50$ & $ 31.4$ &   0   & N & LGG160 &  5  & HG90   &  3   &  0.25     & H97 \\   
NGC 2699      & $-18.72$ & $12.97$ & $ 21.8$ &   4   &   & LGG164 &  4  & ...    & ...  &  ...       & \\   
NGC 2778      & $-19.06$ & $13.06$ & $ 26.6$ &   0   & Y & LGG171 &  3  & GH43   &  5   &  0.42     & G93 \\   
NGC 2824      & $-19.75$ & $13.92$ & $ 54.2$ &   3   &   & ...    &  1  & ...    &  1   &   0        & \\   
NGC 2872      & $-20.38$ & $12.54$ & $ 38.3$ &   4   &   & ...    & ..  & H99    &  2   &  0.0105    & \\  
NGC 2902      & $-18.97$ & $12.86$ & $ 23.3$ &   2   &   & LGG174 &  3  & ...    & ...  &  ...       & \\   
NGC 2950      & $-19.93$ & $11.52$ & $ 19.5$ &   0   & N & ...    &  1  & ...    &  1   &   0       & H97 \\   
NGC 2986      & $-20.87$ & $11.27$ & $ 26.8$ &   0   & N & ...    & ..  & ...    &  2   &  ...      & G94 \\   
NGC 3065      & $-19.56$ & $12.69$ & $ 28.2$ &   4   & Y & ...    & ..  & ...    &  2   &  ...      & BB87, R91\\
NGC 3078      & $-20.71$ & $11.60$ & $ 29.0$ &   4   & Y & LGG185 &  9  & HG29   &  4   &  1.11     & TD91 \\   
NGC 3193      & $-19.62$ & $11.64$ & $ 17.9$ &   0   & L2: & LGG194 & 13  & GH58   &  10  &  0.20   & H95, H97 \\   
NGC 3226      & $-19.11$ & $12.08$ & $ 17.3$ &   3   & L1.9 & LGG194 & 13  & GH58  &  10  &  0.20   & H95, H97 \\       
NGC 3266      & $-19.11$ & $13.18$ & $ 28.6$ &   0   &   & ...    & ..  & GH63   &  4   &  0.41      & \\   
NGC 3348      & $-21.23$ & $11.69$ & $ 38.5$ &   0   & Y & LGG224 &  3  & GH69   &  3   &  0.67     & H95 \\   
NGC 3377      & $-18.89$ & $10.90$ & $  9.1$ &   1   & Y & LGG217 &  9  & GH68   &  23  &  0.47     & G93, B00 \\   
NGC 3414      & $-19.64$ & $11.73$ & $ 18.8$ &   1   & L2 & LGG227 &  6  & GH71  &  7  &  0.34    & H95, H97 \\       
NGC 3595      & $-19.69$ & $12.72$ & $ 30.4$ &   0   &   & ...    & ..  & ...    &  2   &  ...       & \\   
NGC 3613      & $-20.75$ & $11.49$ & $ 28.0$ &   0   & N & LGG232 &  4  & GH94   & 170  &  0.92     & G94, H97\\   
NGC 3640      & $-20.01$ & $11.03$ & $ 16.2$ &   0   & N & LGG233 &  6  & GH76   &  7   &  0.23     & H97 \\   
NGC 4121      & $-18.31$ & $13.95$ & $ 28.3$ &   0   &   & ...    &  1  & ...    &   1  &  0         & \\   
NGC 4125      & $-21.20$ & $10.31$ & $ 20.1$ &   4   & T2 & LGG274 &  3  & GH94   & 170  &  0.92    & H95, H97 \\   
NGC 4128      & $-19.97$ & $12.58$ & $ 32.4$ &   0   &   & LGG272 &  3  & GH99   &  14  &  1.11      & \\   
NGC 4168      & $-20.52$ & $11.83$ & $ 29.5$ &   2   & S1.9: & LGG285 & 25  & GH106  & 248  &  1.08 & R91, H95, H97 \\   
NGC 4233      & $-19.71$ & $12.64$ & $ 29.6$ &   4   & Y & LGG278 &  8  & GH106  & 248  &  1.08     & KF89 \\ 
NGC 4291      & $-20.00$ & $12.04$ & $ 25.6$ &   0   & N & LGG284 & 11  & GH107  &  13  &  0.70     & H97 \\   
NGC 4365      & $-20.77$ & $10.21$ & $ 15.7$ &   0   & N & LGG289 & 63  & GH106  & 248  &  1.08     & S91, TD91, H97 \\
NGC 4474      & $-19.42$ & $12.13$ & $ 20.4$ &   0   &   & LGG289 & 63  & GH106  & 248  &  1.08      & \\   
NGC 4478      & $-19.36$ & $11.93$ & $ 18.1$ &   0   & N & LGG289 & 63  & GH106  & 248  &  1.08     & G93, H97 \\   
NGC 4482      & $-18.52$ & $13.34$ & $ 23.6$ &   0   &   & LGG296 & 11  & GH106  & 248  &  1.08      & \\   
NGC 4494      & $-20.84$ & $10.41$ & $ 17.8$ &   4   & L2:: & LGG294 &  3  & GH94   & 170  &  0.92  & H95, H97 \\       
NGC 4503      & $-19.44$ & $11.79$ & $ 17.6$ &   0   & N & LGG289 & 63  & GH106  & 248  &  1.08     & H97 \\   
NGC 4564      & $-19.15$ & $11.67$ & $ 14.6$ &   0   & N & LGG289 & 63  & GH106  & 248  &  1.08     & G94, H97 \\   
NGC 4589      & $-20.88$ & $11.33$ & $ 27.6$ &   2   & L2 & LGG284 & 11  & GH107  &  13  &  0.70    & G94, H95, H97 \\
NGC 4621      & $-18.50$ & $10.41$ & $  6.0$ &   0   & N & ...    & ..  & GH106  & 248  &  1.08     & H97 \\   
NGC 4648      & $-19.09$ & $12.60$ & $ 21.8$ &   4   & N & LGG303 &  7  & GH107  &  13  &  0.70     & H97 \\   
NGC 5017      & $-19.19$ & $13.25$ & $ 30.8$ &   4   &   & LGG338 &  9  & ...    & ...  &  ...       &  \\   
NGC 5077      & $-20.63$ & $12.03$ & $ 34.0$ &   2   & L1.9 & LGG343 &  3  & ...    & ...  &  ...   & M96, H95, H97 \\
NGC 5173      & $-19.52$ & $13.07$ & $ 32.8$ &   3   & Y & LGG352 &  4  & H99    &  2   &  0.156    & BB87 \\ 
NGC 5198      & $-20.24$ & $12.42$ & $ 34.1$ &   0   & Y & LGG352 &  4  & H99    &  2   &  0.156    & R91 \\   
NGC 5283      & $-18.93$ & $14.13$ & $ 40.8$ &   3   & S2 & ...    &  1  & ...    &   1  &  0       & HB92 \\ 
NGC 5308      & $-20.13$ & $11.99$ & $ 26.6$ &   0   & N & LGG360 &  8  & GH122  &  13  &  0.63     & H97 \\   
NGC 5370      & $-19.45$ & $13.63$ & $ 41.3$ &   0   &   & ...    & ..  & GH122  &  13  &  0.63     &  \\   
NGC 5557      & $-21.52$ & $11.62$ & $ 42.5$ &   0   & N & LGG378 &  3  & GH141  &  13  &  0.02     & H97 \\   
NGC 5576      & $-20.00$ & $11.41$ & $ 19.1$ &   0   & N & LGG379 &  6  & GH139  &  6   &  0.12     & H97 \\   
NGC 5796      & $-20.38$ & $12.43$ & $ 36.5$ &   1   & N & LGG390 &  3  & ...    & ...  &  ...      & R91 \\   
NGC 5812      & $-20.10$ & $11.86$ & $ 24.6$ &   4   & Y & ...    & ..  & H99    &  2   &  0.209    & G93, M96 \\   
NGC 5813      & $-20.85$ & $11.11$ & $ 24.6$ &   4   & L2: & LGG393 &  8  & GH150  &  11  &  0.39   & G93, H95, H97 \\   
NGC 5831      & $-19.59$ & $12.06$ & $ 21.4$ &   0   & Y & LGG393 &  8  & GH150  &  11  &  0.39     & G93, M96 \\   
NGC 5846      & $-21.10$ & $10.74$ & $ 23.3$ &   3   & Y & LGG393 &  8  & GH150  &  11  &  0.39     & BB87, B93, G93 \\
NGC 5898      & $-20.30$ & $11.80$ & $ 26.3$ &   1   & Y & LGG398 &  5  & ...    & ...  &  ...      & R91, M96 \\   
NGC 5903      & $-20.85$ & $11.58$ & $ 30.5$ &   1   & Y & LGG398 &  5  & ...    & ...  &  ...      & M96 \\   
NGC 5982      & $-21.32$ & $11.65$ & $ 39.3$ &   0   & L2:: & LGG402 &  4  & GH158  &  5   &  0.29  & H95, H97 \\   
NGC 6278      & $-19.70$ & $13.15$ & $ 37.1$ &   0   &   & LGG409 &  3  & ...    & ...  &  ...      &  \\   
UGC 4551      & $-18.77$ & $13.09$ & $ 23.6$ &   0   &   & ...    &  1  & ...    &   1  &  0        &  \\   
UGC 4587      & $-19.54$ & $13.49$ & $ 40.3$ &   1   &   & ...    &  1  & ...    &   1  &  0        &  \\   
UGC 6062      & $-18.82$ & $13.74$ & $ 32.7$ &   0   &   & ...    &  1  & ...    &   1  &  0        &  \\      
\enddata

\tablecomments{Column (1): Name of the galaxy. \\ Column (2): Absolute $B$-band magnitude, from Paper I. \\
Column (3): Total apparent $B$-band magnitude, from Paper I. \\ Column (4): Distance in Mpc, from paper I. \\
Column (5): Dust level: 0 = No dust, 1 = Filamentary low, 2 = Filamentary medium, 3 = Filamentary high, 4 = dusty
disk. \\ Column (6): Presence of line emission: Empty entry denotes no observation or no information from literature;
N = no line emission detected, Y = emission line detected. Wherever the type of activity has been determined by 
\citet{ho97a}, this information is given (LINER: L1.9, L2; Transition object: T2; Seyfert: S1.9, S2). Colons denote
uncertainty. \\ Column (7): Lyon Group of Galaxies (LGG) group 
number from \citet{gar93}. \\ Column (8): Number of members in LGG group; = 1 if appeared isolated. \\ Column (9): 
CfA group number from \citet{gh83} and \citet{hg82}, or H99 denoting that the galaxy is a member of a galaxy pair 
in \citet{hon99}. \\ Column (10): Number of members in CfA group, or if = 2, indicating that the galaxy is in an 
interacting pair or having another companion; = 1 if appeared isolated. \\ Column (11): Mean harmonic
radius of the group in Mpc from \citet{gh83} and \citet{hg82}; if a binary, the separation between the pair. \\ 
Column (12): References for emission lines: 
B93=\citet{b93}; B00=\citet{b00}; BB87=\citet{bb87}; BM98=\citet{bm98}; G93=\citet{g93}; G94=\citet{gou94}; 
H95=\citet{ho95}; H97=\citet{ho97a}; HB92=\citet{hb92}; KF89=\citet{kf89}; OD85=\citet{od85}; M96=\citet{mac96}; 
R91=\citet{r91}; S91=\citet{s91}; TD91=\citet{td91}.}

\end{deluxetable}

\clearpage
\begin{deluxetable}{ccccccccc}
\tabletypesize{\scriptsize}
\tablecaption{Snapshot Sample: Dust Properties \label{jdust}}
\tablewidth{0pt}
\tablehead{
\colhead{Name} & \colhead{Morph} & \colhead{$d$} & \colhead{$i$} & 
\colhead{$\langle A_V\rangle$} & \colhead{$M_d$(opt)} & \colhead{$PA_g$} & 
\colhead{$PA_{d}$} & \colhead{$|\Delta PA|$} \\
\colhead{(1)} & \colhead{(2)} & \colhead{(3)} & \colhead{(4)} & \colhead{(5)} & 
\colhead{(6)} & \colhead{(7)} & \colhead{(8)} & \colhead{(9)}
}
\startdata
ESO 437-15 & F3  &     &           & 0.17~ & 3.70 &  41  &  42 &   1 \\
ESO 580-26 & F3  &     &           & 0.21~ & 4.42 &  41  &  30 &  11 \\
NGC 2592   & D   & 0.6 & 34        & 0.093 & 1.53 &  50  &  55 &   5 \\
NGC 2699   & D   & 0.6 & 40        & 0.029 & 0.96 &  45  &  45 &   0 \\
NGC 2824   & F3  &     &           & 0.16~ & 5.00 & $-20$  & 145 &  15 \\
NGC 2872   & D   & 0.9 & 63        & 0.20~ & 2.15 & $-30$  & 155 &   5 \\
NGC 2902   & F2  &     &           & 0.061 & 2.06 &  10  & 112 &  78 \\
NGC 3065   & D   & 0.5 & \nodata   & 0.10~ & 1.28 & $-60$  & \nodata & \nodata \\
NGC 3078   & D   & 1.4 & 75        & 0.33~ & 2.50 & $-01$  & 175 &   4 \\
NGC 3226   & F3  &     &           & 0.13~ & 3.35 &  43  &  15 &  28 \\
NGC 3377   & F1  &     &           & 0.047 & 2.13 &  41  &  50 &   9 \\
NGC 3414   & F1  &     &           & 0.034 & 1.78 &  01  &  93 &  88 \\
NGC 4125   & D   & 3.3 & $\sim$ 90 & 0.11~ & 3.90 & $-98$  &  72 &  10 \\
NGC 4168   & F2  &     &           & 0.035 & 2.22 & $-52$  & 105 &  23 \\
NGC 4233   & D   & 5.4 & $\sim$ 90 & 0.52~ & 4.42 &  25  &  77 &  52 \\
NGC 4494   & D   & 1.6 & 50        & 0.20~ & 1.78 &  01  &  10 &   9 \\
NGC 4589   & F2  &     &           & 0.073 & 4.00 & $-88$  &  00 &  88 \\
NGC 4648   & D   & 0.4 & \nodata   & 0.10~ & 0.98 &  65  & \nodata & \nodata \\
NGC 5017   & D   & 0.5 & $\sim$ 0  & 0.029 & 1.18 &  32  & \nodata & \nodata \\
NGC 5077   & F2  &     &           & 0.048 & 3.96 &  09  & 102 &  87 \\
NGC 5173   & F3  &     &           & 0.055 & 3.68 & $-79$  & 115 &  14 \\
NGC 5283   & F3  &     &           & 0.173 & 4.70 & $-85$  & 109 &  14 \\
NGC 5796   & F1  &     &           & 0.067 & 3.67 & $-76$  &  16 &  88 \\
NGC 5812   & D   & 0.4 & $\sim$ 10 & 0.11~ & 1.43 &  60  & \nodata & \nodata \\
NGC 5813   & D   & 1.4 & $\sim$ 65 & 0.083 & 3.70 & $-39$  & 150 &   9 \\
NGC 5846   & F3  &     &           & 0.030 & 3.41 & $-101$ &  52 &  27 \\
NGC 5898   & F1  &     &           & 0.035 & 2.74 & $-70$  &  00 &  70 \\
NGC 5903   & F1  &     &           & 0.063 & 3.16 & $-14$  & $-04$ &  10 \\
UGC 4587   & F1  &     &           & 0.029 & 2.02 &  07  &  05 &   2 \\
\enddata

\tablecomments{Column (1): Galaxy name. \\ Column (2): Morphology of the dust features: F = filaments, 
with a number denoting the level of dust as in Paper I; D = dusty disk. \\ Column (3): Diameter of the 
dusty disk in arcsec. \\ Column (4): Inclination of the dusty disk in degrees, derived from the 
ellipticity and the assumption that the disk is intrinsically circular. \\ Column (5): Mean visual extinction of 
the dust features in magnitude. \\ Column (6): Dust mass in units of log $M_\sun$ derived from visual 
extinction of the visible dust features. \\ Column (7): PA of the major axis of the galaxy measured 
at 10\arcsec. \\ Column (8): PA of the major axis of the main dust structure. \\ Column (9): Absolute 
value of the difference $PA_g - PA_d$. 
}

\end{deluxetable}

\begin{deluxetable}{lccc}
\tabletypesize{\scriptsize}
\tablecaption{Dust Mass Comparison to Previous Studies \label{mdcomp}}
\tablewidth{0pt}
\tablehead{
\colhead{Galaxy} & \colhead{This work} & \colhead{vD\&F95\tablenotemark{a}} & \colhead{Tom00\tablenotemark{b}} 
}
\startdata   
NGC 3377 &    4.2   &  \nodata &   4.8 \\
NGC 4494 &    3.9   &     4.0  &   5.0 \\
NGC 4589 &    6.1   &     5.8  &   7.0 \\
NGC 5813 &    5.8   &     4.8  &   6.5 \\
\enddata

\tablecomments{The table lists gas mass in logarithmic solar mass units. Previous values were
converted using distances adopted in this paper listed in Table \ref{gentab}.
A gas-to-dust ratio of 130 has been assumed.}

\tablenotetext{a}{\citet{dkfr95}}
\tablenotetext{b}{\citet{tom00}}

\end{deluxetable}

\clearpage
\begin{deluxetable}{cccccccccc}
\tabletypesize{\scriptsize}
\tablecaption{Snapshot Sample: Radio and Infrared Properties \label{jrir}}
\tablewidth{0pt}
\tablehead{
\colhead{} & \colhead{$S_{1400}$} & \colhead{$\sigma_{1400}$} &
\colhead{$S_{60}$} & \colhead{$\sigma_{60}$} & \colhead{$S_{100}$} & 
\colhead{$\sigma_{100}$} & \colhead{$S_{60}/S_{100}$} & \colhead{$T_d$\tablenotemark{a}} & 
\colhead{$M_d(IRAS)$\tablenotemark{b}}
}
\startdata   
ESO 437-15 & 0.0035  & 0.0006  & 0.83 & 0.068 & 1.16 & 0.094 & 0.72 & 44.5 & 5.18 \\
ESO 447-30 & \nodata & \nodata & 0.30 & 0.042 & 1.44 & 0.100 & 0.21 & 27   & 6.27  \\   
ESO 580-26 & 0.0037  & 0.0007  & 0.98 & 0.074 & 1.83 & 0.110 & 0.54 & 39.5 & 5.73 \\
NGC 2549 & \nodata & \nodata & 0.26 & 0.048 & 0.37 & 0.131 & 0.70 & 44   & 4.07 \\
NGC 2634\tablenotemark{c} & \nodata & \nodata & 0.28 & 0.032 & 0.98 & 0.165 & 0.29 & 30   & 5.77 \\
NGC 2778 & \nodata & \nodata & 0.00 & 0.041 & 0.51 & 0.096 & 0.00 & 30   & 5.34 \\
NGC 2824 & 0.0093  & 0.0005  & 0.94 & 0.074 & 1.51 & 0.100 & 0.62 & 42   & 5.83 \\
NGC 2872 & 0.0074  & 0.0026  &      &       &      &       &      &      &      \\
NGC 2902 & \nodata & \nodata & 0.16 & 0.040 & 1.04 & 0.183 & 0.15 & 24   & 6.06 \\
NGC 2950 & \nodata & \nodata & 0.16 & 0.033 & 0.20 & 0.115 & 0.80 & 48   & 3.87 \\
NGC 2986 & \nodata & \nodata & 0.00 & 0.025 & 0.40 & 0.123 & 0.00 & 30   & 5.25 \\
NGC 3065\tablenotemark{c} & 0.0045  & 0.0006  & 1.51 & 0.025 & 2.00 & 0.100 & 0.76 & 46   & 5.25 \\
NGC 3078 & 0.3138  & 0.0109  &      &       &      &       &      &      &      \\
NGC 3226 & 0.0033  & 0.0005  &      &       &      &       &      &      &      \\
NGC 3348 & 0.0083  & 0.0005  & 0.13 & 0.029 & 0.30 & 0.139 & 0.43 & 35   & 5.14 \\
NGC 3377 & \nodata & \nodata & 0.14 & 0.045 & 0.35 & 0.065 & 0.40 & 34.5 & 3.98 \\
NGC 3414 & 0.0047  & 0.0005  & 0.25 & 0.025 & 0.56 & 0.185 & 0.45 & 36   & 4.73 \\
NGC 4125 & \nodata & \nodata & 0.70 & 0.044 & 1.67 & 0.069 & 0.42 & 35   & 5.32 \\
NGC 4168 & 0.0060  & 0.0015  & 0.00 & 0.036 & 0.66 & 0.157 & 0.00 & 30   & 5.55 \\
NGC 4233 & 0.0034  & 0.0006  & 0.19 & 0.037 & 0.48 & 0.087 & 0.40 & 34.5 & 5.14 \\
NGC 4365 & \nodata & \nodata & 0.00 & 0.044 & 0.65 & 0.131 & 0.00 & 30   & 4.99 \\
NGC 4494 & \nodata & \nodata & 0.19 & 0.049 & 0.00 & 0.170 & .... & 30   & 4.90 \\
NGC 4564 & $\lesssim$ 0.0025 & &     &       &      &       &      &      &      \\ 
NGC 4589 & 0.0378  & 0.0015  & 0.20 & 0.031 & 0.66 & 0.153 & 0.30 & 31   & 5.42 \\
NGC 5077 & 0.1608  & 0.0057  &      &       &      &       &      &      &      \\
NGC 5173 & 0.0032  & 0.0006  & 0.35 & 0.042 & 0.53 & 0.159 & 0.66 & 43.5 & 4.88 \\
NGC 5198 & 0.0040  & 0.0006  &      &       &      &       &      &      &      \\
NGC 5283 & 0.0134  & 0.0004  &      &       &      &       &      &      &      \\
NGC 5576 & \nodata & \nodata & 0.09 & 0.027 & 0.21 & 0.278 & 0.43 & 35   & 4.37 \\
NGC 5796 & 0.1100  & 0.0044  &      &       &      &       &      &      &      \\
NGC 5812 & $\lesssim$ 0.002~ & &     &       &      &       &      &      &      \\ 
NGC 5813 & 0.0158  & 0.0011  &      &       &      &       &      &      &      \\
NGC 5846 & 0.0221  & 0.0014  &      &       &      &       &      &      &      \\
NGC 5898 & \nodata & \nodata & 0.13 & 0.036 & 0.23 & 0.072 & 0.56 & 40   & 4.46 \\
NGC 5903 & 0.0318  & 0.0021  &      &       &      &       &      &      &      \\
NGC 5982 & \nodata & \nodata & 0.00 & 0.033 & 0.37 & 0.035 & 0.00 & 30   & 5.54 \\
NGC 6278 & \nodata & \nodata & 0.00 & 0.023 & 0.32 & 0.100 & 0.00 & 30   & 5.43 \\
UGC 4587 & \nodata & \nodata & 0.14 & 0.028 & 0.94 & 0.061 & 0.15 & 24   & 6.50 \\
\enddata

\tablecomments{$S_{1400}$ ($\sigma_{1400}$), $S_{60}$ ($\sigma_{60}$), $S_{100}$ ($\sigma_{100}$)
are the 1400 MHz, 60$\mu$m and 100$\mu$m flux densities (and their associated errors) in Jy, respectively.} 
\tablenotetext{a}{Dust temperature in Kelvin, derived from color ratio $S_{60}/S_{100}$. When 
$S_{60}/S_{100}$ is not available, a temperature of 30K is assumed.}
\tablenotetext{b}{Dust mass in units of log $M_\sun$ derived from $IRAS$ flux densities and dust color 
temperature.}
\tablenotetext{c}{Infrared fluxes may likely be contaminated with emission from nearby sources \citep{knap89}.}

\end{deluxetable}

\begin{deluxetable}{lcrl}
\tabletypesize{\scriptsize}
\tablecaption{Pearson Correlation Coefficients \label{rtab}}
\tablewidth{0pt}
\tablehead{
\colhead{Relationship} & \colhead{$N$} & \colhead{$r$} & \colhead{$p$}  
}
\startdata
log $L_{IR60}$ -- log $M_d$(opt)  & 13\tablenotemark{a}~ & 0.76\tablenotemark{a}~ & 0.0032\tablenotemark{a} \\
                              & 10 & 0.80~ & 0.0075 \\
log $L_{IR100}$ -- log $M_d$(opt) & 13\tablenotemark{a} & 0.57\tablenotemark{a}~ & 0.044\tablenotemark{a}~ \\
                              & 11 & 0.63~ & 0.040~ \\
log $L_{rad}$ -- log $M_d$(opt)   & 18 & 0.21~ & 0.41~~ \\
                              & 13 & 0.43~ & 0.15~~ \\
log $L_{IR60}$ -- $M_B$           & 20 & 0.14~ & 0.54~~ \\
                              & 16 & 0.10~ & 0.71~~ \\
log $L_{IR100}$ -- $M_B$          & 25 & 0.080 & 0.70~~ \\
                              & 22 & 0.094 & 0.68~~ \\
log $L_{rad}$ -- $M_B$            & 20 & $-$0.44 & 0.051 \\
                              & 15 & $-$0.42 & 0.12~ \\
$M_B$ -- log $M_d$(opt)       & 29 & $-$0.15 & 0.44~ \\
                              & 11\tablenotemark{b} & $-$0.83\tablenotemark{b} & 0.003\tablenotemark{b} \\
\enddata

\tablecomments{For each pair of variables, the number of data points ($N$), the 
linear correlation coefficient ($r$), and the probability of the null 
hypothesis (no correlation, $p$) are shown in the first line for all galaxies 
with dusty disks and filamentary dust. The second line shows the results for 
galaxies with filamentary dust only, except for $M_B$ -- log $M_d$(opt), where 
it shows the coefficients for galaxies with dusty disks.}
\tablenotetext{a}{Excluding NGC 3065}
\tablenotetext{b}{Excluding NGC 4233}

\end{deluxetable}

\clearpage
\begin{deluxetable}{lccccc}
\tabletypesize{\scriptsize}
\tablecaption{Detection Rates \label{drate}}
\tablewidth{0pt}
\tablehead{
\colhead{} & \colhead{Dusty Disks (12, 15)} & \colhead{Fil Dust (17, 16)} & \colhead{All Dust (29, 31)} & 
\colhead{No Dust (38, ~9)} & \colhead{All Gal (67, 40)}
}
\startdata
EL -- snapshot & ~8~~7~~~88 & 14~12~86 & 22~19~86 &  21~~6~28 &  43~25~58 \\
~~~~ -- $IRAS$  & 14~12~~~86 & 16~12~75 & 30~24~80 &  ~8~~2~25 &  38~26~68 \\
        &           &          &          &           &           \\
Radio -- snapshot & 12~~6~~~50 & 17~13~76 & 29~19~66 &  38~~3~~8 &  67~22~33 \\
~~~~~~~~ -- $IRAS$   & 13~~8~~~62 & 16~~8~50 & 29~16~55 &  ~9~~2~22 &  38~18~47 \\
        &           &          &          &           &           \\
Far-IR\tablenotemark{a}~ -- snapshot  & 10~~4~~~40 & 15~11~73 & 25~15~60 &  37~11~30 &  62~26~42 \\
        &           &          &          &           &           \\
Group -- snapshot  & 12~12~100 & 16~13~81 & 28~25~89 &  36~31~86 &  64~56~88 \\
\enddata

\tablecomments{Detection rate of emission lines (EL), radio emission from NVSS, 
far-IR emission from $IRAS$ survey, and being in a group/cluster/binary pair. 
For each group, the first number indicates the total number of galaxies 
with available data, the second number denotes the number of galaxies with 
detections, and the last number gives the percentage relative to 
the number of galaxies with available data within that group. 
Values for the snapshot sample are given in the first line, and for the $IRAS$ sample in the second line.
The first and second number in parentheses in the heading give the total number of galaxies
in each group for the snapshot and $IRAS$ sample, respectively.}

\tablenotetext{a}{By definition, the far-IR detection rate for the $IRAS$ sample is 100\%.}

\end{deluxetable}

\begin{deluxetable}{lcccccl}
\tabletypesize{\scriptsize}
\tablecaption{$IRAS$ Sample \label{irastab}}
\tablewidth{0pt}
\tablehead{
\colhead{Name} & \colhead{$v$}  & \colhead{$B_T$} & \colhead{Dust} & \colhead{Radio} & \colhead{Line Emiss} & \colhead{Ref} \\
\colhead{(1)} & \colhead{(2)} & \colhead{(3)} & \colhead{(4)} & \colhead{(5)} & \colhead{(6)} & \colhead{(7)} \\
}
\startdata	  
 ESO 208-G021    & 1037  &    12.19  &  4   &    \nodata &    Y  &    P86 \\
 ESO 358-G059    & 1007  &    13.99  &  0   &    N       &       &        \\
 IC 1459         & 1691  &    10.97  &  4   &    Y       &    Y  &    B93, R91, G94 \\
 NGC 205(M110)   & $-$241  &     8.92  &  1   &    N       &    N  &    H97 \\
 NGC 404         & $-$48  &    11.21  &  3   &    Y       &   L2  &    R91, H97 \\
 NGC 584         & 1875  &    11.44  &  1   &    N       &    Y  &    M96 \\
 NGC 821         & 1718  &    11.67  &  0   &    N       &    N  &    G93, H97 \\
 NGC 1052        & 1470  &    11.41  &  2   &    Y       &  L1.9 &    H97 \\
 NGC 1339        & 1392  &    12.51  &  0   &    N       &    N  &    R91 \\
 NGC 1351        & 1511  &    12.46  &  0   &    N       &    N  &    P86 \\
 NGC 1399        & 1425  &    10.55  &  0   &    Y       &    Y  &    G94, M96 \\
 NGC 1400        &  558  &    11.92  &  2   &    Y       &    N  &    M96 \\
 NGC 1404        & 1947  &    11.92  &  1   &    Y       &    N  &    G94 \\
 NGC 1439        & 1670  &    12.27  &  4   &    N       &    N  &    R91 \\
 NGC 2768        & 1339  &    10.84  &  4   &    Y       &   L2  &    H97 \\
 NGC 2778\tablenotemark{a}       & 2032  &    13.35  &  0   &    N       &    Y  &    G93 \\
 NGC 2974        & 2072  &    11.87  &  3   &    Y       &    Y  &    R91 \\
 NGC 2986\tablenotemark{a}       & 2329  &    11.72  &  0   &    N       &    N  &    G94 \\
 NGC 3156        & 1118  &    13.07  &  2   &    N       &    Y  &    BB87 \\
 NGC 3377\tablenotemark{a}       &  692  &    11.24  &  1   &    N       &    Y  &    G93, B00 \\
 NGC 3610        & 1787  &    11.70  &  3   &    N       &    Y  &    G94 \\
 NGC 4125\tablenotemark{a}       & 1356  &    10.65  &  4   &    N       &   T2  &    H97 \\
 NGC 4168\tablenotemark{a}       & 2284  &    12.11  &  2   &    Y       &  S1.9 &    R91, H97 \\
 NGC 4261        & 2210  &    11.14  &  4   &    Y       &   L2  &    G93, H97 \\
 NGC 4278        &  649  &    11.09  &  3   &    Y       &  L1.9 &    R91, H97 \\
 NGC 4365\tablenotemark{a}       & 1240  &    10.52  &  0   &    N       &    N  &    S91, TD91, H97 \\
 NGC 4374(M84)   & 1000  &    10.09  &  4   &    Y       &    L2 &    H97 \\
 NGC 4406(M86)   & $-$227  &     9.83  &  4   &    N       &    N  &    H97 \\
 NGC 4476        & 1978  &    13.01  &  3   &    N       &    N  &    R91 \\
 NGC 4486(M87)   & 1282  &     9.59  &  4   &    Y       &    L2 &    H97 \\
 NGC 4552(M89)   &  321  &    10.73  &  4   &    Y       &   T2: &    H97  \\
 NGC 4589\tablenotemark{a}       & 1980  &    11.69  &  2   &    Y       &    L2 &    G94, H97 \\
 NGC 4649(M60)   & 1413  &     9.81  &  0   &    Y       &    N  &    H97 \\
 NGC 4697        & 1236  &    10.14  &  4   &    N       &    Y  &    G93 \\
 NGC 4742        & 1270  &    12.12  &  3   &    N       &    Y  &    R91 \\
 NGC 5128(Cen A) &  547  &     7.84  &  4   &    Y       &    Y  &    P86  \\
 NGC 5322        & 1915  &    11.14  &  4   &    Y       &  L2:: &    H97 \\
 NGC 5845        & 1450  &    13.50  &  4   &    N       &       &        \\ 
 NGC 5898\tablenotemark{a}       & 2209  &    12.49  &  1   &    N       &    Y  &    R91, M96 \\
 NGC 6861        & 2819  &    12.12  &  4   &    \nodata &    Y  &    P86    \\ 
\enddata

\tablecomments{Column (1): Name of the galaxy. \\ Column (2): Heliocentric velocity in \kms. \\ Column (3): 
Total apparent $B$-band magnitude. \\ Column (4): Dust level as in Table \ref{gentab}. \\ Column (5): 
Radio detection at 1.4 Ghz from NVSS or FIRST survey. \\ Column (6): Presence of line emission. \\ Column (7): 
References for emission lines as in Table \ref{gentab}; also P86=\citet{p86}.}

\tablenotetext{a}{Also in the snapshot sample.}
  
\end{deluxetable}


\begin{thebibliography}{}
\bibitem[Barnes \& Hernquist(1996)]{bh96} Barnes, J. E., \& Hernquist, L. 1996, \apj, 471, 115
\bibitem[Barnes \& Hernquist(1998)]{bh98} Barnes, J. E., \& Hernquist, L. 1998, \apj, 495, 187
\bibitem[Bennett \& Moss(1998)]{bm98} Bennett, S. M., \&  Moss, C. 1998, \aaps, 132, 55
\bibitem[Bertola \& Corsini(1999)]{bc99} Bertola, F., \& Corsini, E. M. 1999, in {\it IAU Symp. 186: 
  Galaxy Interactions at Low and High Redshift}, ed. J. E. Barnes \& D. B. Sanders, 186, 149
\bibitem[Bettoni \& Buson(1987)]{bb87} Bettoni, D., \& Buson, L. M. 1987, \aaps, 67, 341
\bibitem[Bower et al.(1998)]{bow98} Bower, G. A. et al. 1998, \apj, 492, L111
\bibitem[Bureau et al.(2000)]{b00} Bureau, M. et al. 2000, in {\it Galaxy Disks and Disk Galaxies}, 
  ASP Conference Series, ed. J. G. Funes S. J. \& E. M. Corsini, astro-ph/0009332
\bibitem[Buson et al.(1993)]{b93} Buson, L. M., Sadler, E. M., Zeilinger, W. W., Bertin, G., Bertola,
  F., Danzinger, J., Dejonghe, H., Saglia, R. P., \& de Zeeuw, P. T. 1993, \aap, 280, 409 
\bibitem[Caldwell(1984)]{c84} Caldwell, N. 1984, \pasp, 96, 287
\bibitem[Calvani et al.(1989)]{cff89} Calvani, M., Fasano, G., \& Franceschini, A. 1989, \aj, 97, 1319 
\bibitem[Caon, Macchetto \& Pastoriza(2000)]{c00} Caon, N., Macchetto, D. \& Pastoriza, M. 2000, \apjs, 
 127, 39 
\bibitem[Cardelli et al.(1989)]{ccm89} Cardelli, J. A., Clayton, G. C., \& Mathis, J. S. 1989, \apj, 
  345, 245
\bibitem[Condon et al.(1998)]{con98} Condon, J. J., Cotton, W. D., Greisen, E. W., Yin, Q. F., Perley,
  R. A., Taylor, G. B., \& Broderick, J. J. 1998, \aj, 115, 1693
\bibitem[Davies et al.(1987)]{dav87} Davies, R. L., Burstein, D., Dressler, A., Faber, S. M., 
  Lynden-Bell, D., Terlevich, R. J., \& Wegner, G. 1987, \apjs, 64, 581
\bibitem[de Koff et al.(2000)]{dkof00} de Koff, S., Best, P., Baum, S. A., Sparks, W.,
  Rottgering, H., Miley, G., Golombek, D., Maccheto, F., \& Martel, A. 2000, \apjs, 129, 33
\bibitem[Faber \& Jackson(1976)]{fj76} Faber, S. M., \& Jackson, R. E. 1976, \apj, 668, 683
\bibitem[Faber et al.(1987)]{f87} Faber, S. M. et al. 1987, in {\it Nearly Normal Galaxies}, 
  ed. S. Faber, (New York: Springer), p. 175
\bibitem[Faber et al.(1997)]{fab97} Faber, S. M. et al. 1997, \aj, 114, 1771 
\bibitem[Ferrarese \& Ford(1999)]{ff99} Ferrarese, L., \& Ford, H. C. 1999, \apj, 515, 583
\bibitem[Ferrarese et al.(1996)]{ffj96} Ferrarese, L., Ford, H. C., \& Jaffe, W. 1996, \apj, 470, 444
\bibitem[Ferrarese \& Merritt(2000)]{fm00} Ferrarese, L., \& Merritt, D. 2000, \apj, 539, L9
\bibitem[Ferrarese et al.(1994)]{f94} Ferrarese, L., van den Bosch, F. C., Ford, H. C., Jaffe, W., \& 
  O'Connell, R. W.\ 1994, \aj, 108, 1598
\bibitem[Ferrari et al.(1999)]{fer99} Ferrari, F., Pastoriza, M. G., Macchetto, F., \& Caon, N. 1999,
  \aaps, 136, 269
\bibitem[Forbes(1991)]{f91} Forbes, D. A. 1991, \mnras, 249, 779
\bibitem[Ford et al.(1997)]{f97} Ford, H. C., Tsvetanov, Z. I., Ferrarese, L., Kriss, G., Jaffe, 
  W., Harms, R. \& Dressel, L. 1997, in {\it Accretion Phenomena and Related Outflows}, IAU Coll. 
  No. 163, ed. D. T. Wickramasinghe, G. V. Bicknell, and L. Ferrario, p. 620
\bibitem[Garcia(1993)]{gar93} Garcia, A. M. 1993, \aaps, 100, 47
\bibitem[Gebhardt et al.(2000)]{g00} Gebhardt, K. et al. 2000, \apj, 539, L13
\bibitem[Geller \& Huchra(1983)]{gh83} Geller, M. J., \& Huchra, J. P. 1983, \apjs, 52, 61
\bibitem[Goudfrooij et al.(1994)]{gou94} Goudfrooij, P., Hansen, L., J\o rgensen, H. E., \& 
  N\o rgaard-Nielsen, H. U. 1994, \aaps, 105, 341
\bibitem[Goudfrooij \& de Jong(1995)]{gdj95} Goudfrooij, P., \& de Jong, T. 1995, \aap, 298, 784
\bibitem[Gonzalez(1993)]{g93} Gonzalez, J. J. 1993, Ph.D. Thesis, University of California, Santa Cruz
\bibitem[Harms et al.(1994)]{h94} Harms, R. J. et al. 1994, \apj, 435, L35
\bibitem[Ho et al.(1995)]{ho95} Ho, L. C., Filippenko, A. V., \& Sargent, W. L. W. 1995, \apjs, 98,
  477 
\bibitem[Ho et al.(1997a)]{ho97a} Ho, L. C., Filippenko, A. V., \& Sargent, W. L. W. 1997a, \apjs, 
  112, 315
\bibitem[Ho et al.(1997b)]{ho97b} Ho, L. C., Filippenko, A. V., \& Sargent, W. L. W. 1997b, \apj, 487,
  568
\bibitem[Honma(1999)]{hon99} Honma, M. 1999, \apj, 516, 693
\bibitem[Huchra \& Burg(1992)]{hb92} Huchra, J. P., \& Burg, R. 1992, \apj, 393, 90
\bibitem[Huchra \& Geller(1982)]{hg82} Huchra, J. P., \& Geller, M. J. 1982, \apj, 257, 423
\bibitem[Jaffe et al.(1994)]{j94} Jaffe, W., Ford, H. C., O'Connell, R. W., van den Bosch, F. C., \& 
  Ferrarese, L. 1994, \aj, 108, 1567 
\bibitem[Kannappan \& Fabricant(2000)]{kf00} Kannappan, S. J., \& Fabricant, D. G. 2000, \aj, 121, 140
\bibitem[Kim(1989)]{k89} Kim, D.-W. 1989, \apj, 346, 653
\bibitem[Knapp et al.(1989)]{knap89} Knapp, G.\ R., Guhathakurta, P., Kim, D.\ and Jura, M.\ A.\ 1989,
  \apjs, 70, 329
\bibitem[Kollatschny \& Fricke(1989)]{kf89} Kollatschny, W., \& Fricke, K. J. 1989, \aap, 219, 3
\bibitem[Kormendy \& Richstone(1995)]{kr95} Kormendy, J., \& Richstone, D. O. 1995, \araa, 33, 581
\bibitem[Kwan \& Xie(1992)]{kx92} Kwan, J., \& Xie, S. 1992, \apj, 398, 105
\bibitem[Lauer et al.(1995)]{l95} Lauer, T. R. et al. 1995, \aj, 110, 2622 
\bibitem[Macchetto et al.(1997)]{mac97} Macchetto, F., Marconi, A., Axon, D. J., Capetti, A., 
  Sparks, W. \& Crane, P. 1997, \apj, 489, 579 
\bibitem[Macchetto et al.(1996)]{mac96} Macchetto, F., Pastoriza, M., Caon, N., Sparks, W. B., 
  Giavalisco, M., Bender, R., \& Capaccioli, M. 1996, \aaps, 120, 463 
\bibitem[Martel et al.(2000)]{mar00} Martel, A. R., Turner, N. J., Sparks, W. B., \& Baum, S. A. 2000,
  \apjs, 130, 267 
\bibitem[Martini \& Pogge(1999)]{mp99} Martini, P., \& Pogge, R. W. 1999, \aj, 118, 2646
\bibitem[McElroy(1995)]{mcel95} McElroy, D. B. 1995, \apjs, 100, 105
\bibitem[Merluzzi(1998)]{mer98} Merluzzi, P. 1998, \aap, 338, 807
\bibitem[Neilsen \& Tsvetanov(2000)]{nt00} Neilsen, E. H., \& Tsvetanov, Z. I. 2000, \apj, 536, 255  
\bibitem[Phillips et al.(1986)]{p86} Phillips, M. M., Jenkins, C. R., Dopita, M. A., Sadler, E. M.,
  \& Binette, L. 1986, \aj, 91, 1062
\bibitem[Osterbrock \& de Robertis(1985)]{od85} Osterbrock, D. E., \& de Robertis, M. M. 1985, \pasp, 
  97, 1129
\bibitem[Rest et al.(2001)]{rbos00} Rest, A., van den Bosch, F. C., Jaffe, W., Tran,
  H. D., Tsvetanov, Z., Ford, H. C., Davies, J., \& Schafer, J. 2001, \aj, in press, astro-ph/0102286 (Paper I)
\bibitem[Roberts et al.(1991)]{r91} Roberts, M. S., Hogg, D. E., Bregman, J. N., Forman, W. R., \&
  Jones, C. 1991, \apjs, 75, 751
\bibitem[Sadler \& Gerhard (1985)]{sg85} Sadler, E. M., \& Gerhard, O. E. 1985, \mnras,
  214, 177
\bibitem[Shields(1991)]{s91} Shields, J. C. 1991, \aj, 102, 1314
\bibitem[Sparks et al.(2000)]{s00} Sparks, W. B., Baum, S. A., Biretta, J., Macchetto, D., \& Martel, 
  A. R. 2000, \apj, 542, 667
\bibitem[Steiman-Cameron \& Durisen(1988)]{scd88} Steiman-Cameron, T. Y., \& Durisen, R. H. 1988, 
  \apj, 325, 26
\bibitem[Tohline et al.(1982)]{tsc82} Tohline, J. E., Simonson, G. F., \& Caldwell, 
  N. 1982, \apj, 252, 92
\bibitem[Tomita et al.(2000)]{tom00} Tomita, A., Aoki, K., Watanabe, M., Takata, T., 
  \& Ichikawa, S. 2000, \aj, 120, 123
\bibitem[Trinchieri \& di Serego Alighieri(1991)]{td91} Trinchieri, G., \& di Serego Alighieri, S.
  1991, \aj, 101, 1647
\bibitem[Tsai \& Mathews(1996)]{tm96} Tsai, J. C., \& Mathews, W. G. 1996, \apj, 468, 571
\bibitem[van den Bosch et al.(1994)]{vdb94} van den Bosch, F. C., Ferrarese, L., Jaffe, W., Ford, 
  H. C., \& O'Connell, R. W. 1994, \aj, 108, 1579 
\bibitem[van der Marel(1999)]{vdm99} van der Marel, R. P. 1999, in {\it Galaxy Interactions at Low 
  and High Redshifts}, IAU Symp. No. 186, ed. D. B. Sanders, \& J. Barnes, (Kluwer:Dordrecht), p. 333
\bibitem[van der Marel \& van den Bosch(1998)]{vv98} van der Marel, R. P. \& van den Bosch, F. 1998, 
  AJ, 116, 2220
\bibitem[van Dokkum \& Franx(1995)]{dkfr95} Van Dokkum, P. G., \& Franx, M. 1995, \aj, 110, 2027
\bibitem[Verdoes Kleijn et al.(1999)]{vk99} Verdoes Kleijn, G. A., Baum, S. A., de Zeeuw, P. T., 
  \& O'Dea, C. P. 1999, \aj, 118, 2592
\bibitem[Verdoes Kleijn et al.(2000)]{vk00} Verdoes Kleijn, G. A., van der Marel, R. P., Carollo, 
  C. M., \& de Zeeuw, P. T. 2000, \aj, 120, 1221
\bibitem[Verkhodanov et al.(1997)]{v97} Verkhodanov, O. V., Trushkin S. A., Andernach H., \& Chernenkov,
  V. N. 1997, in {\it Astronomical Data Analysis Software and Systems VI}, ASP Conf. Ser., ed. G. Hunt
  \& H. E. Payne, 125, 322
\bibitem[V\'{e}ron-Cetty \& V\'{e}ron(1988)]{vv88} V\'{e}ron-Cetty, M.-P., \& V\'{e}ron, P. 1988,
  \aap, 204, 28
\end{thebibliography}
\end{document}